\documentclass[12pt,preprint]{aastex}
\newcommand{\om}{\Omega_{\rm M}}
\newcommand{\ola}{\Omega_\Lambda}
\newcommand{\ogal}{\Omega_{\rm G}}

\shorttitle{Corrections for lensing of supernovae}

\begin{document}
\title{CORRECTIONS FOR GRAVITATIONAL LENSING OF SUPERNOVAE: BETTER THAN AVERAGE?}
\author{
  C.~Gunnarsson, T.~Dahl\'en, A.~Goobar, and J.~J\"onsson}
\affil{Stockholm University, AlbaNova University Center, \\
  Fysikum, SE-10691 Stockholm, Sweden}

\and

\author{E.~M\"ortsell}
\affil{Stockholm University, AlbaNova University Center, \\
  Stockholm Observatory, SE-10691 Stockholm, Sweden}
\email{
  cg@physto.se
}

\begin{abstract}
We investigate the possibility of correcting for the magnification due
to gravitational lensing of standard candle sources, such as Type Ia
supernovae. Our method uses the observed properties of the foreground
galaxies along the lines-of-sight to each source and the accuracy of
the lensing correction depends on the quality and depth of these
observations as well as the uncertainties in translating the observed
luminosities to the matter distribution in the lensing galaxies. The
current work is limited to cases where the matter density is
dominated by the individual galaxy halos. However, it is
straightforward to generalize the method to include also gravitational
lensing from cluster scale halos. We show that the dispersion due to
lensing for a standard candle source at $z=1.5$ can be reduced from
about $7\,\%$ to $\lesssim 3\,\%$,~i.e.~the magnification correction
is useful in reducing the scatter in the Type Ia Hubble diagram,
especially at high redshifts where the required long exposure times
makes it hard to reach large statistics and the dispersion due to
lensing becomes comparable to the intrinsic Type Ia scatter.
\end{abstract}


\section{INTRODUCTION}\label{sec:intro}

Observations of high-redshift Type Ia supernovae (SNIa) have in the
last decade or so
lead to a dramatic paradigm shift in cosmology
\citep{perl98,riess98,schmidt,perl99,knop03,tonry,riess04}.
Measurements of the luminosity distance to supernovae over a wide
range of redshifts were used to break the degeneracy between cosmic
fluids, as suggested by \citet{goo95}. The data clearly favors a
universe dominated by repulsive dark energy, and which is presently undergoing
accelerated expansion. The next step in observational cosmology is to
test the nature of this dark energy, whether constant, i.e.~compatible with
Einstein's cosmological constant, or due to completely new physics.
Observations of SNIa are among the leading astrophysical tools to
explore this question further, as they probe the expansion history of
the Universe directly. Large dedicated surveys are in progress
(e.g. CFHTLS, ESSENCE, SDSSII) and even more ambitious space based projects are
being planned for the future, e.g. the JDEM proposals, DESTINY, JEDI and SNAP.

One thing in common for all these projects is the very large projected number of
SNIa that eventually will populate the Hubble diagram used to derive
cosmological parameters. Clearly, systematic uncertainties will (soon)
become the limiting factor. While some of these uncertainties are due
to our lack of knowledge of the SNIa physics and intrinsic properties,
others stem from possible interactions of the supernova light
(rest-frame UV and optical) near the source or along the
line-of-sight (l-o-s), e.g. extinction by dust in the host galaxy or
intergalactic medium. In this work, we focus on the gravitational
interaction of photons along the l-o-s, i.e.~gravitational
lensing. As supernova surveys become deeper, the measured source
fluxes become increasingly more sensitive to the inhomogeneities in
the matter distribution of the Universe.  
In \citet{ramme}, the
JDEM/SNAP mission was simulated using the SNOC Monte-Carlo package
\citep{snoc} and it was found that a careful {\em statistical}
treatment could be used to 
optimize the fitting of cosmological parameters from the
Hubble diagram of SNIa taking into account the (asymmetric) redshift dependent 
lensing magnification distribution. Lensing on {\em
individual} SNe have also been studied, in e.g
\citet{lewibata,97ff,benitez,qlet}
by modeling the effect from the galaxies close to the
l-o-s to the SN.  

In this work, we investigate the accuracy to which lensing
(de)magnification can be estimated on individual supernovae. For that
purpose, we create mock galaxy catalogs with properties (e.g., galaxy 
magnitudes, redshifts and spectral types) based on luminosity functions 
derived from observations by \citet{dah05}. Using the brightness of 
galaxies as a tracer of the gravitational fields along the l-o-s, we use the
multiple lens-plane package Q-LET
\citep{qlet} to investigate the accuracy to which the magnification
can be estimated as a function of the survey parameters, assumptions
on M/L-ratios and halo shapes. In an accompanying paper \citep{jacke},
we apply the technique described here to investigate the lensing
magnification probability distribution for 33 supernovae in the 
GOODS survey \citep{riess04,str04}.

In this paper, we will assume that dark matter halos in individual
galaxy halos and small groups are most important for the lensing
magnification of supernovae. For cluster size lenses, additional
information is needed to model the gravitational potential, e.g.,
lensing of background galaxies. Though such a generalization of the
method is straightforward, in the following we have assumed the
uncertainty in the lensing magnification factor for the small fraction
of SNe with foreground clusters to be of the same order as SNe with
massive foreground galaxies. Also, we assume that the large scale dark
matter structures in the Universe is traced by the luminous matter,
i.e.~that filaments and walls are populated by galaxies. This approach
is reasonable as long as the luminous and dark matter are not {\em
anti-}correlated, see \S\ref{sec:summary}.

In \S\ref{sec:correct}, we discuss whether one needs to correct for
lensing at all. Section \S\ref{sec:modeling} describes the underlying
theory and \S\ref{sec:simsurv} treats our method for estimating the
accuracy of the lensing corrections. We summarize and discuss our
results in \S\ref{sec:summary}. Throughout the paper we use natural
units, where $c=G=1$. We assume that the underlying cosmological
parameters are $H_0=70\ {\rm km}\ \!{\rm s}^{-1} {\rm Mpc}^{-1}$, the
matter density $\om=0.3$ and the dark energy density $\ola=0.7$. When
no explicit redshift dependence is shown, quantities refer to present
values ($z=0$). Quoted magnitudes are in the Vega system.

\section{TO CORRECT OR NOT TO CORRECT}\label{sec:correct}
Since the mean magnification due to gravitational lensing of a large
number of sources is expected to be unity relative to a homogeneous
universe, the question arises whether one should correct for the
effect of gravitational lensing at all.

Because flux, $f$, is conserved, it is the mean of the magnification {\em factor},
$\mu$, that is equal to one, i.e.~$\bar\mu =1$, or defining $\mu =1+\delta$ where
$\delta$ is the fractional difference in luminosity from the unlensed
(homogeneous universe) case, $\bar \delta=0$. The magnitude is given by
$m=-2.5\log f + {\rm const}$, and we can write
\begin{equation}
	m = m_0 -\frac{2.5}{\ln 10}\ln\mu ,
\end{equation}
where $m_0$ is the unlensed magnitude. Taylor expanding $\ln\mu =\ln
(1+\delta)$, we get
\begin{equation}
	m = m_0 -\frac{2.5}{\ln 10}\left[\delta-\frac{\delta^2}{2}+{\mathcal O}(\delta^3)\right],
\end{equation}
with mean value $\bar m =m_0+0.54\bar{\delta^2} + {\mathcal
O}(\bar{\delta^3})$. From this it is clear that the average lensed
magnitude need not be equal to the unlensed magnitude. Note that in
current surveys, this effect is very small compared to, e.g., the
intrinsic scatter of SN luminosities which is why the distinction is
still unimportant. Note also that the mean magnification factor is
unity only for random source positions. For an actual sample of
observed SNIa, magnification bias can push the mean magnification to
higher values.
 
However, given that we have a sample of random source position SNIa
and neglect the small corrections to $\bar m$ (or perform our
cosmology fit using flux units), then $\bar m$ is an unbiased
estimator for the population mean of the observed
magnitudes\footnote{As long as the variance of $m$ is finite.}. Under these
circumstances, neglecting the scatter due to lensing does not
cause any bias in the fitted cosmological parameters and good
statistics will help in beating down the error
\citep[e.g.,][]{holz04}. There could still be good reasons to consider
correcting for lensing effects. If we are able to reduce the scatter
in the observed magnitudes and keep $\bar m$ as an unbiased estimator,
then we are able to make more accurate cosmology fits. There are also
cases where it is non-trivial to quantify the importance of the
magnification bias, e.g., the case of SN 1997ff. In a similar context,
the ability to correct individual lines-of-sight for gravitational
lensing magnification would have a profound impact in our ability use
gravitational wave ``sirens'' for measuring cosmological parameters,
as their use as standard candles is ultimately limited by the lensing
uncertainty \citep{holz&hughes}.

\section{MODELING AN INHOMOGENEOUS UNIVERSE}\label{sec:modeling}
In this section we present a method to investigate the effects of
gravitational lensing in an inhomogeneous universe.

\subsection{Halo Profiles}\label{subs:halo}
Neglecting gravitational lensing is equivalent to assuming that matter
is homogeneously distributed in the Universe. However, on small scales,
the Universe is certainly inhomogeneous. To investigate the
effects of gravitational lensing of distant sources, a realistic
model of the matter distribution in the Universe is needed. In the
following, we describe how we (re)distribute the matter in our model
universe using observations of the luminous matter.

We assume that each galaxy is surrounded by a dark matter halo and
that the mass of this halo can be estimated from the galaxy
luminosity. However, inferring masses of dark matter halos from luminosities
of galaxies is non-trivial. The effects of lensing by a halo
depends not only on its mass, but also on its density profile. Both
the density profile and mass of dark matter halos are issues under
debate. We have chosen to   
work mainly with two different halo models,
Singular Isothermal Spheres (SIS) and the model of Navarro, Frenk and
White \citep[NFW; ][]{nfwref}.

The density profile of a SIS, $\rho_{\rm SIS}(r)=\sigma^2/(2 \pi
r^2)$, is characterized by its l-o-s velocity dispersion $\sigma$,
which can be estimated from the galaxy luminosity via the Faber--Jackson
(F--J) or Tully--Fisher (T--F) relations, approximately valid for
elliptical and spiral galaxies respectively. Since the mass of a SIS
halo diverges, $m_{\rm SIS}(r)=2 \sigma^2 r$, we use a truncation
radius $r_{\rm t}$. A commonly used scale for halo profiles in general
is $r_{\rm 200}$, defined as the radius inside which the mean mass
density is 200 times the present critical density. For a SIS halo, $r_{\rm
200}$ and the corresponding mass within this radius, $m_{\rm 200}$,
are given by
\begin{equation}
r_{\rm 200}^{\rm SIS}=\frac{\sqrt{2} \sigma}{10H_0},\;m_{\rm 200}^{\rm
SIS}=\frac{\sqrt{2}\sigma^3}{5H_0}.
\end{equation}
The density profile of a NFW halo is
\begin{equation}
\rho_{\rm NFW}(r)=\frac{\rho_{\rm s}}
  {(r/r_{\rm s})(1+r/r_{\rm s})^2}, 
\end{equation}
where $r_{\rm s}$ is the scale radius for which approximately
$\rho_{\rm NFW}\propto r^{-2}$ 
and $\rho_{\rm s}$ is the density at $r\sim 0.5 r_{\rm s}$.

The NFW halo is fully determined by $m_{\rm 200}$ since $r_{\rm
200}=\left[ m_{\rm 200}/(100 H_0^2)\right]^{1/3}$ and the scale radius
$r_{\rm s}$ and $\rho_{\rm s}$ can be found numerically from $m_{\rm
  200}$ \citep{nfwref}.  

In the following, we assume that the mass within $r_{\rm 200}$ is
roughly the same for the SIS and NFW halo profiles, i.e.~$m_{\rm
200}^{\rm SIS}=m_{\rm 2 00}^{\rm NFW}$. We also set the truncation
radius $r_{\rm t}=r_{\rm 200}$ for both SIS and NFW halos. Varying
$r_{\rm t}$ does not alter gravitational lensing effects
significantly, see \S\ref{subs:halochoice}.

\subsection{The Smoothness Parameter}\label{subs:smooth}

Very faint and/or small scale structures cannot all be seen in a
magnitude limited survey. Also, for the method used in this paper, any
matter not directly associated with individual galaxies such as
completely dark halos and cluster halos need to be accounted for. In
order to assure that the mean mass density in our model universe is
kept constant, we keep the ``remaining'' mass, not accounted for when
relating the dark matter to the luminous matter, as a homogeneous
component. 

The homogeneous part can be characterized by the
\emph{smoothness parameter} $\eta(z)$, quantifying the fraction of
smoothly distributed matter in our model universe (or our lack of
knowledge on the dark matter distribution in the real Universe). Since
the fraction of galaxies observed at a given magnitude limit is a
function of redshift, and also since the Universe evolves, the
smoothness parameter is expected to vary with redshift.

The smoothness parameter in a given survey can be computed from the
{\em observed} density of matter in clumps, i.e.~in our case galaxies
surrounded by dark matter halos, $\rho_{\rm g}(z)$. If the redshift
dependence of $\rho_{\rm g}(z)$ can be factorized into a term
$(1+z)^3$, scaling like the matter density, and an unknown factor
$f(z)$ originating from the magnitude limit of the survey and
evolution, we can write
\begin{equation}
\rho_{\rm g}(z)=\rho_{\rm g}(0)(1+z)^3 f(z).
\end{equation}
Then the smoothness parameter is
simply given by
\begin{equation}
\eta(z)=1-\frac{\ogal}{\om}f(z),
\end{equation}
where the density in galaxies at $z=0$ has been scaled with the
present critical density to $\ogal$.
Once the galaxies have been associated with halos of definite masses,
the comoving density of clumps as a function of redshift $\ogal f(z)$
can be estimated. We divide the distribution of galaxies into redshift
bins and estimate $\ogal f(z)$ in each bin. The density of clumps in
the $i$:th bin, centered on redshift $z_i$, is obtained through
\begin{equation}
 \ogal f(z_i)=\frac{1}{\rho_{\rm c}}\frac{\sum_j m_j}{V_i},
\end{equation}
where $m_j$ is the mass of a clump and $\rho_{\rm c}$ is the critical
density. The comoving volume of the $i$:th 
bin is given by
\begin{equation}
V_i=\int_{z_i-\Delta z/2}^{z_i+\Delta z/2}  
\frac{D_{\rm A}^2 (1+z)^2}{\left[ \om (1+z)^3 +\ola \right]^{1/2}}
{\Delta\Omega} dz,
\label{eq:vol}
\end{equation}
where $\Delta z$ is the width of the bin, $\Delta \Omega$ is the solid
angle under study and $D_{\rm A}$ is the angular diameter distance.
Distances have been calculated using the \texttt{angsiz} routines
described in \citet{helbig}, in which a smoothness parameter varying with redshift
can be included. Note that the angular diameter distance $D_{\rm A}$ used to determine the
volume element in Eq.~(\ref{eq:vol}) above is calculated using the
filled-beam approximation ($\eta=1$), since the volume is governed by
the global expansion rate, which in turn is governed by the properties
on very large scales where the Universe is homogeneous.

\subsection{Deriving Velocity Dispersions from Observed Luminosities}\label{subs:vdfromlum}
Galaxy halo masses can be estimated from the velocity dispersion of
galaxies. We calculate the velocity dispersion of each galaxy using
absolute magnitudes ($M_B$) combined with empirical F--J and T--F
relations. For ellipticals,
we use the F--J relation
\begin{equation}
\log_{10}\sigma =\log_{10}\sigma_*-\frac{0.4}{\gamma}(M_B-M_B^*)
\label{eq:FJ}
\end{equation}
where $M_B^*$ is the characteristic magnitude and $\sigma_*$ is the
normalization in velocity dispersion.  We use $\gamma=4.4$, as derived
by \citet{mit05} using Sloan Digital Sky Survey (SDSS) data. To derive $\sigma_*$, we use
equation~(33) in \citet{mit05}, where we use the relation
$M_r=M_B-1.32$ to convert SDSS $r$--band magnitudes in AB system to 
standard $B$--band Vega normalized magnitudes. We have here assumed a typical
color $M_B-M_r=1.20$~for ellipticals in the AB-system, and an AB to Vega 
relation B$_{\rm AB}$=B$_{\rm Vega}-0.12$. The normalization in 
velocity dispersion is
given by
\begin{equation}
\log_{10}\sigma_*=2.2-0.091(M_B^*+19.47+0.85z),
\label{eq:star}
\end{equation}
where we use $M_B^*=-21.04$ derived for the early--type population
by \citet{dah05}. Equation~(\ref{eq:star}) yields $\sigma_*=220$
${\rm km \, s^{-1}}$ at $z=0$. Combining equation~(\ref{eq:FJ})
and~(\ref{eq:star}) gives an expression for the velocity dispersion
\begin{equation}
\log_{10}\sigma=-0.091(M_B-4.74+0.85z'),
\label{eq:sigma}
\end{equation}
where we use $z'=z$ for redshifts $z<1$ and $z'=1$ for $z> 1$. The
redshift dependence of the relation accounts for the brightening of
the stellar population with redshift. Since this evolution is poorly
known at $z>1$, we assume a flat evolution at these redshifts. As a
measurement of the error in the derived relation, we use the observed
scatter in the SDSS measurements by \citet{she03}
\begin{equation}
{\rm rms}(\log_{10}\sigma)=0.079[1+0.17(M_B+19.705+0.85z')],
\label{eq:rms}
\end{equation}
where we again have transformed SDSS $r$--band to standard $B$--band
magnitudes. 

For the spiral and later type population, we use the T--F relation derived
by \citet{pie92}, with correction for redshift calculated
by \citet{boh04}
\begin{equation}
\log_{10}V_{\rm max}=-0.134(M_B-\Delta M_B+3.52),
\label{eq:TF}
\end{equation}
where $V_{\rm max}$ is the maximum rotation velocity for the galaxy.
The correction due to redshift dependence is
\begin{equation}
\Delta M_B=-1.22z'-0.09.
\end{equation}
The observed scatter in the relation derived by \citet{pie92} is ${\rm
rms}(M_B)=0.41$, corresponding to
\begin{equation}
{\rm rms}(\log_{10}V_{\rm max})=0.06.
\label{eq:scatter}
\end{equation}
At $M_B^*$, this is similar to the errors in the F--J relation
above. Finally, the velocity dispersion in spiral galaxies is related
to the circular velocity via $\sigma=V_{\rm max}/\sqrt{2}$.

\subsection{Gravitational Lensing with Multiple Lenses}\label{subs:lensing}
A typical source l-o-s within some angular
radius $\theta_{\rm s}$ will contain more than
one lens. This requires the multiple lens-plane algorithm
\citep[see][for further details]{
schneider,qlet}, which takes into
account each lens along the l-o-s by projecting the lens' mass
distribution onto a plane and then traces the light-ray from the image
plane (first lens-plane) back through all lens-planes up to the source
plane where the magnification and intrinsic position can be found.  

In the following we denote angular diameter distances between
redshifts $z_i$ and $z_j$ by $D_{ij}$. We use $o$ for observer, $s$
for source and $d$ for lens (deflector). When $z_i=0$, that index is
omitted. 

Each halo is truncated in 3D at $r=r_{\rm t}$, then, upon projection
onto a plane, the corresponding surface mass density will smoothly go
to zero at the projected truncation radius. The projection can be done
analytically for our lens models. For simplicity, we start by
considering a single lens-plane. The equations can be simplified if we
let $\xi$ be the impact parameter on a halo and define $x=\xi/\xi_0$
and $x_{\rm t}=r_{\rm t}/\xi_0$, where
\begin{equation} \xi_0=\frac{4\pi \sigma^2 D_{\rm d}D_{\rm ds}}{D_{\rm
s}} \label{eq:xsizerosis} \end{equation} for the SIS and
\begin{equation} \xi_0=r_{\rm s} \label{eq:xsizeronfw}
\end{equation} for the NFW halo. Then, the projected density $\kappa
(x)$ can be written as
\begin{equation}
\kappa_{\rm SIS}(x)=\frac{1}{\pi x} \arctan \left(
\frac{\sqrt{x_{\rm t}^2-x^2}}{x} \right)
\label{eq:sisproj}
\end{equation}
and
\begin{equation}
\kappa_{\rm NFW}(x)=\frac{2\kappa_{\rm s}}{x^2-1}f(x),
\label{eq:nfwproj1}
\end{equation}
where
\begin{equation}
f(x)=\left\{ \begin{array}{ccc}
 \frac{\sqrt{x_{\rm t}^2-x^2}}{1+x_{\rm t}}+\frac{1}{\sqrt{1-x^2}}\left( {\rm arctanh}
 \left( \sqrt{\frac{x_{\rm t}^2-x^2}{1-x^2}}\right)-{\rm arctanh}\left(\frac{
 \sqrt{\frac{x_{\rm t}^2-x^2}{1-x^2}}}{x_{\rm t}}\right)\right) & \  & x<1<x_{\rm t} \\
 \frac{\sqrt{x_{\rm t}^2-x^2}}{1+x_{\rm t}}+\frac{1}{\sqrt{x^2-1}}\left( \arctan\left(\frac{
 \sqrt{\frac{x_{\rm t}^2-x^2}{x^2-1}}}{x_{\rm t}}\right)-\arctan
 \left( \sqrt{\frac{x_{\rm t}^2-x^2}{x^2-1}}\right)\right) & \  & 1<x\leq x_{\rm t} \\
\end{array} \right.
\label{eq:fofx}
\end{equation}
and
\begin{equation}
\kappa_{\rm NFW}(1)=\frac{2}{3}\kappa_{\rm s}\left(
\frac{x_{\rm t}^3-3x_{\rm t}+2}{\left( x_{\rm t}^2-1\right)^{\frac{3}{2}}} \right). 
\label{eq:nfwproj2}
\end{equation}
Here, $\kappa_{\rm s}=\rho_{\rm s}r_{\rm s}/\Sigma_{\rm cr}$, where
$\Sigma_{\rm cr}=D_{\rm s}/4\pi D_{\rm d}D_{\rm ds}$ is a critical
density related to strong lensing. Note that $x_{\rm t}>1$ must be
assumed for the NFW and that $\kappa=0$ for $x>x_{\rm t}$ for both
halo types. 

The general expression for the deflection angle for circularly
symmetric lenses is
\begin{equation}
\hat{\alpha}(x)=s\alpha(x)=s\frac{2}{x}\int^x_0 x'\kappa(x')dx',
\label{eq:deflgen}
\end{equation}
where $s=\xi_0 D_{\rm s}/D_{\rm d}D_{\rm ds}$, resulting in
\begin{equation}
\alpha(x)=\frac{2}{\pi}\left(  \arctan \left(
\frac{\sqrt{x_{\rm t}^2-x^2}}{x} \right)+\frac{x_{\rm t}-\sqrt{x_{\rm t}^2-x^2}}{x} \right)
\label{eq:sisdefl}
\end{equation}
for the SIS model. For the NFW halo, numerical evaluation is needed.

As the magnification factor, $\mu'$, for all halo models is obtained
with distances calculated with the $z$-dependent $\eta$ function, this
is the universe relative to which $\mu'$ is found (implying $\mu'\geq
1$ for primary images). In the following, we will quote
magnifications, $\mu$, relative to a universe with homogeneously
distributed matter, the filled-beam value (fb) where $\bar\mu =1$. The
magnifications are related by
\begin{equation}
\mu=\mu' \left(\frac{D_{\rm s}^{\rm fb}}{D_{\rm s}^{\eta({\rm z})}}\right)^2.
\label{eq:mu}
\end{equation}  

\section{SIMULATED SURVEYS}\label{sec:simsurv}
In order to study gravitational lensing corrections, we perform Monte
Carlo simulations where we calculate the magnification factor for
random source positions in mock galaxy catalogs. By varying the
assumptions of the galaxy mass distributions as well as the magnitude
limit of the observations, we can estimate the accuracy to which it is
possible to correct for the lensing magnification.

For all lensing calculations we have used the publicly available
\texttt{fortran 77} code Q-LET\footnote{Available at 
\texttt{http://www.physto.se/\~{}cg/qlet/qlet.htm}} \citep{qlet}, although
substantially modified. The code fully utilizes the multiple
lens-plane algorithm and has been used previously by \citet{qlet} and \citet{riess04}
to study lensing effects on supernovae.

As our simulation base, we create for each Monte Carlo realization a
mock galaxy catalog designed to reflect the distribution of galaxies
expected in a circular cone around a random l-o-s. To characterize the
galaxy population we use the $B$-band rest-frame Schechter luminosity
function (LF) derived by \citet{dah05} using GOODS CDF-S
observations. The LF is used to generate the number of expected
galaxies within the cone where we take into account Poissonian
fluctuations but do not include effects of galaxy correlations or
cosmic variance. The same LF is used to assign absolute magnitudes to
each object within the range $-23<M_B<-16$. To account for
evolutionary effects, we include a brightening of the $B$-band
characteristic magnitude by $\sim$1 mag to redshift $z=1$ as discussed
in \S\ref{subs:vdfromlum}.

A random spectral type is assigned according to the type-specific LF
of early-types, late-types and starburst galaxies at $z\sim$0.4 
in \citet{dah05}. We thereafter assign early-type galaxies an 
elliptical morphology and late-type galaxies a spiral morphology.
We assume that the fraction of galaxies with elliptical morphology 
is constant over the redshift range investigated.

The redshift of each object is assigned with a probability proportional
to the volume element, $dV(z)/dz$, which is equivalent to assuming a
constant comoving number density of galaxies with redshift, i.e.~we do not
include any evolution of the number densities due 
to e.g., mergers or large scale structures. The galaxy is finally
given a random position within the l-o-s cone.

Besides redshift, absolute magnitude, spectral type and position, we
also calculate the apparent magnitude in the observed $I$-band for
each object. This allows us to draw subsamples from the catalog with
imposed magnitude cutoffs as is the case for real
observations. Furthermore, to resemble an observational situation
where redshifts are determined photometrically, we also have the
option to add a random error to the redshifts. These errors are
calculated using simulations and depend on redshift, 
spectral type and detection S/N. For bright 
objects with high S/N, errors are typically
$\Delta_z\equiv\langle|z_{\rm phot}-z_{\rm true}|/(1+z_{\rm spec})\rangle\sim
0.05$,
while errors for faint
objects (mostly at high $z$) can be as large as $\Delta
z\sim 0.3$. The errors for early-type galaxies are typically a factor two
smaller compared to late-type galaxies when comparing at the same
apparent magnitudes. 
Besides this "Gaussian" contribution to the error distribution, a 
fraction of the galaxies may also be assigned "catastrophic 
redshifts" with large errors. We discuss this further 
in \S\ref{subs:reduncert}.

\begin{figure}
\epsscale{0.6}
\plotone{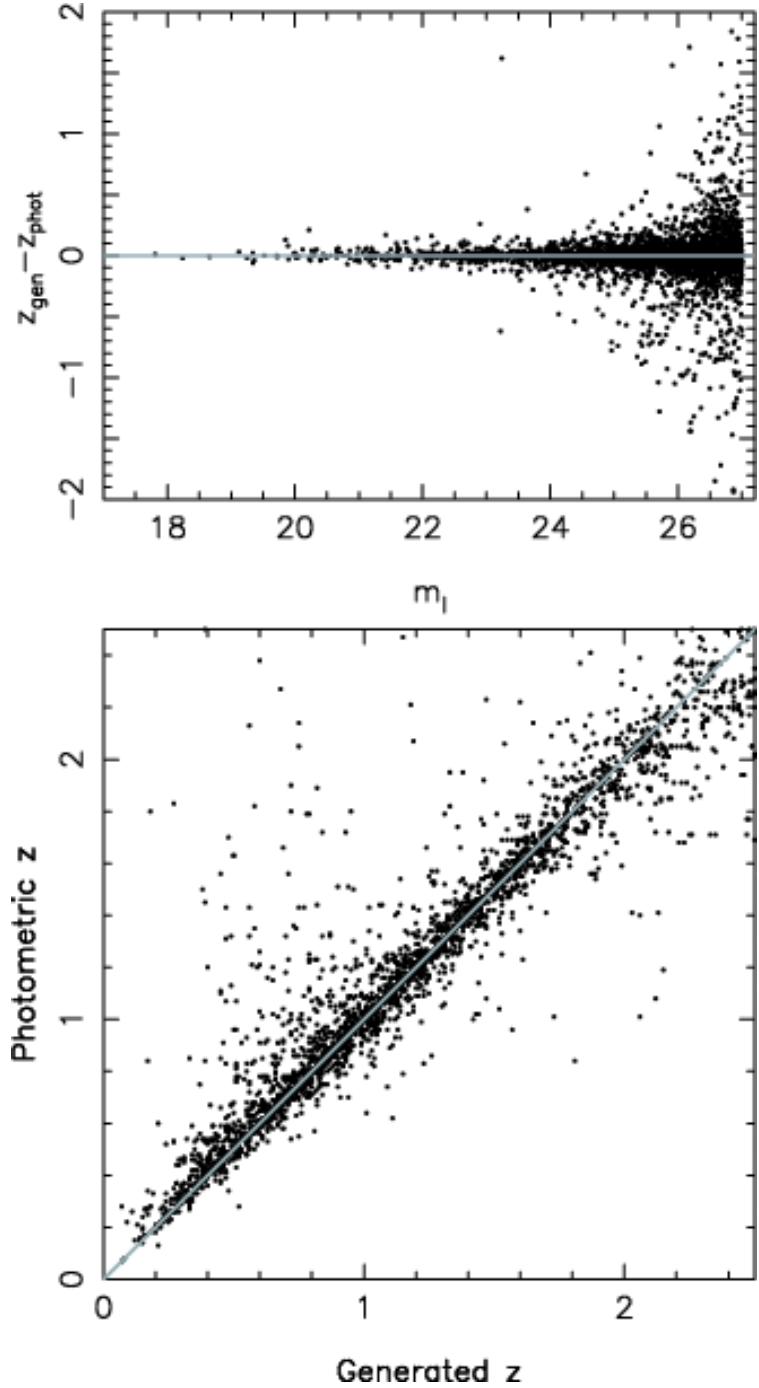}
\caption{\label{fig:photz} Bottom panel shows generated (input) redshifts
vs.~derived photometric redshifts for a simulated survey with limiting 
magnitude $I<27$ (S/N=10). Top panel shows the difference between generated 
and photometric redshifts as a function of observed galaxy magnitude.}
\end{figure}

Figure \ref{fig:photz} shows the simulated accuracy of 
the photometric redshifts for a survey with limiting magnitude $I<27$ (S/N=10).
The bottom panel shows the generated (input) redshifts vs. the photometric
redshifts, while the top panel shows the difference between generated and 
photometric redshifts as a function of galaxy magnitude.

\subsection{Understanding the Magnification Uncertainties}\label{subs:understand}
We have identified and addressed the following uncertainties when
estimating the lensing magnification of a specific source given a
galaxy catalog: 

\begin{itemize}

\item Finite field size

\item The intrinsic scatter in, and accuracy of, the F--J and T--F relations

\item Redshift and position uncertainties

\item Choice of halo profile 

\item The magnitude limit

\end{itemize}

\begin{figure}
\plotone{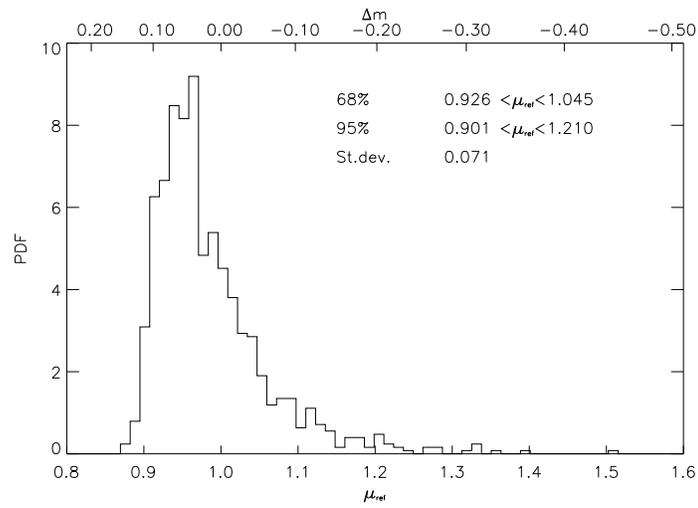}
\caption{\label{fig:default} PDF of magnifications for the 
reference model with NFW halos truncated at $r_{\rm t}=r_{\rm 200}$,
  no magnitude limit, velocity dispersion/circular velocity
  normalizations 220 km$\ \!$s$^{-1}$ for ellipticals, 203 km$\ \!$s$^{-1}$ for spirals, no
  uncertainty in halo redshifts and source redshift $z=1.5$}
\end{figure}

These sources of error are addressed individually in the following
sections. Our reference model consists of NFW halos truncated at
$r_{\rm t}=r_{\rm 200}$, velocity dispersion/circular velocity
normalizations of 220 km$\ \!$s$^{-1}$ for ellipticals, 203 km$\
\!$s$^{-1}$ for spirals and source redshift $z=1.5$. In
Figure~\ref{fig:default}, the Probability Distribution Function (PDF)
for lensing magnifications for the reference model is shown. The
lensing dispersion at $z=1.5$ is $\sim 7\,\%$.  We analyze the results
by comparing the distribution of magnifications in the reference model
with the distribution obtained after performing a correction with the
above-mentioned uncertainties. We denote the uncorrected value
$\mu_{\rm ref}$ and the corrected one $q_{\mu}$, where
\begin{equation}
\label{eq:q_c}
q_{\mu}=\frac{\mu_{\rm ref}}{\mu_{\rm est}},
\end{equation}
where $\mu_{\rm est}$ is the estimated magnification factor including
one or more uncertainties.
Corrections will reduce the uncertainties from lensing whenever the
width of the distribution of $q_{\mu}$ is smaller than the
corresponding width in the $\mu_{\rm ref}$ distribution. If no
uncertainties were present $\mu_{\rm est}=\mu_{\rm ref}$ and $q_{\mu
}=1$ implying a perfect correction. As a measure of the width we give
the standard deviation of the distribution. Since many of the
distributions are non-Gaussian, we also report the 68\,\% and 95\,\%
confidence levels.

\subsection{Finite Field Size}\label{subs:field}
\begin{figure}
\epsscale{0.5}\plotone{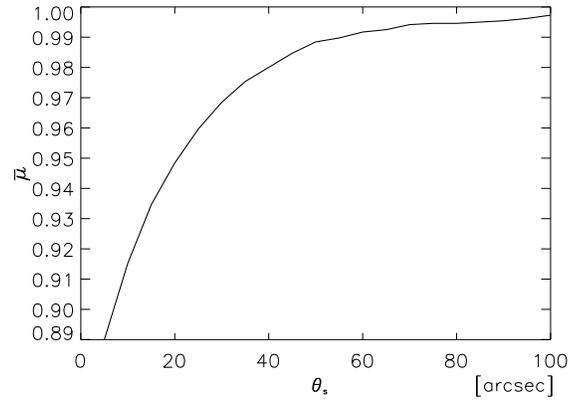}
\caption{\label{fig:rhost} Mean magnification $\bar \mu$ of 5000 lines-of-sight
  as a function of $\theta_{\rm s}$. The sources all have redshifts $z=1.5$.}
\end{figure}

The mean magnification relative to an homogeneous universe of a large
number of sources lensed by randomly distributed matter is expected to be
unity due to photon number conservation. When we model a lensing
system, only galaxies within angular radius $\theta_{\rm s}$ of the
position of the source on the sky are taken into account and thus
$\bar\mu <1$.  If $\theta_{\rm s}$ is increased, more lenses
are added and the mean magnification increases. This dependence is
illustrated in Figure~\ref{fig:rhost}, showing the mean magnification of
5000 point sources, as a function of $\theta_{\rm s}$. The mean
magnification increases rapidly for small $\theta_{\rm s}$, but only
slowly for $\theta_{\rm s}$ larger than an arc-minute, where the error
is $\sim 1\,\%$. In our simulations, we use $60''$ as a cutoff to save
computing time. In a real survey, a cutoff will have to be introduced
for practical purposes since only a limited portion of the sky will be
observed. In order to avoid a systematic bias due to the finite field
size for a given survey, the computed magnifications should be
corrected with a factor corresponding to the inverse of the mean
magnification for the cutoff radius used. In the following, we have
neglected this small ($\sim 1\,\,\%$) correction.
Furthermore, going to larger $\theta_{\rm s}$ would not render
$\bar \mu$ being 
exactly unity since some flux is lost whenever multiple imaging
occurs. Q-LET gives the magnification and intrinsic source position of
a given \emph{image}, not observed position and magnification of a
given source. Therefore, in the rare cases of multiple imaging, only
one of the images will be taken into account resulting in some flux
loss. Note also that since random l-o-s and random source
positions are different \citep[e.g.,][]{schneider}, we have to use a
magnification dependent weighting procedure to see whether each
simulated event should be kept 
or discarded in order to get a sample of random source positions
\citep[see e.g.,][]{snoc}. 

\subsection{The F--J and T--F Relations}\label{subs:fjtf} 
The F--J and T--F relations give the velocity dispersions of the
\emph{luminous} matter and we make the assumption that the dark matter
that constitutes the halo follows scales in the same way. 
Both the Faber-Jackson and the Tully-Fisher relations have an intrinsic
scatter with an rms estimated in \S\ref{subs:vdfromlum}.
To study the effect of this scatter, we add random offsets to the 
F--J and T--F relations when calculating the halo mass. Since we do not want
to bias the total mass in our simulations, we distribute the offsets 
using a Gaussian distribution in $\sigma^3$~(since mass $\propto \sigma^3$). 
We derive the width of the Gaussian (one sigma value in $\sigma^3$)
from the rms in log$_{10}(\sigma)$~and log$_{10}(V_{\rm max})$~using 
Eqs.~(11)-(12) (F--J) and Eqs.~(13)-(15) (T--F).

\begin{figure}
\begin{center}
\resizebox{0.45\textwidth}{!}{\includegraphics{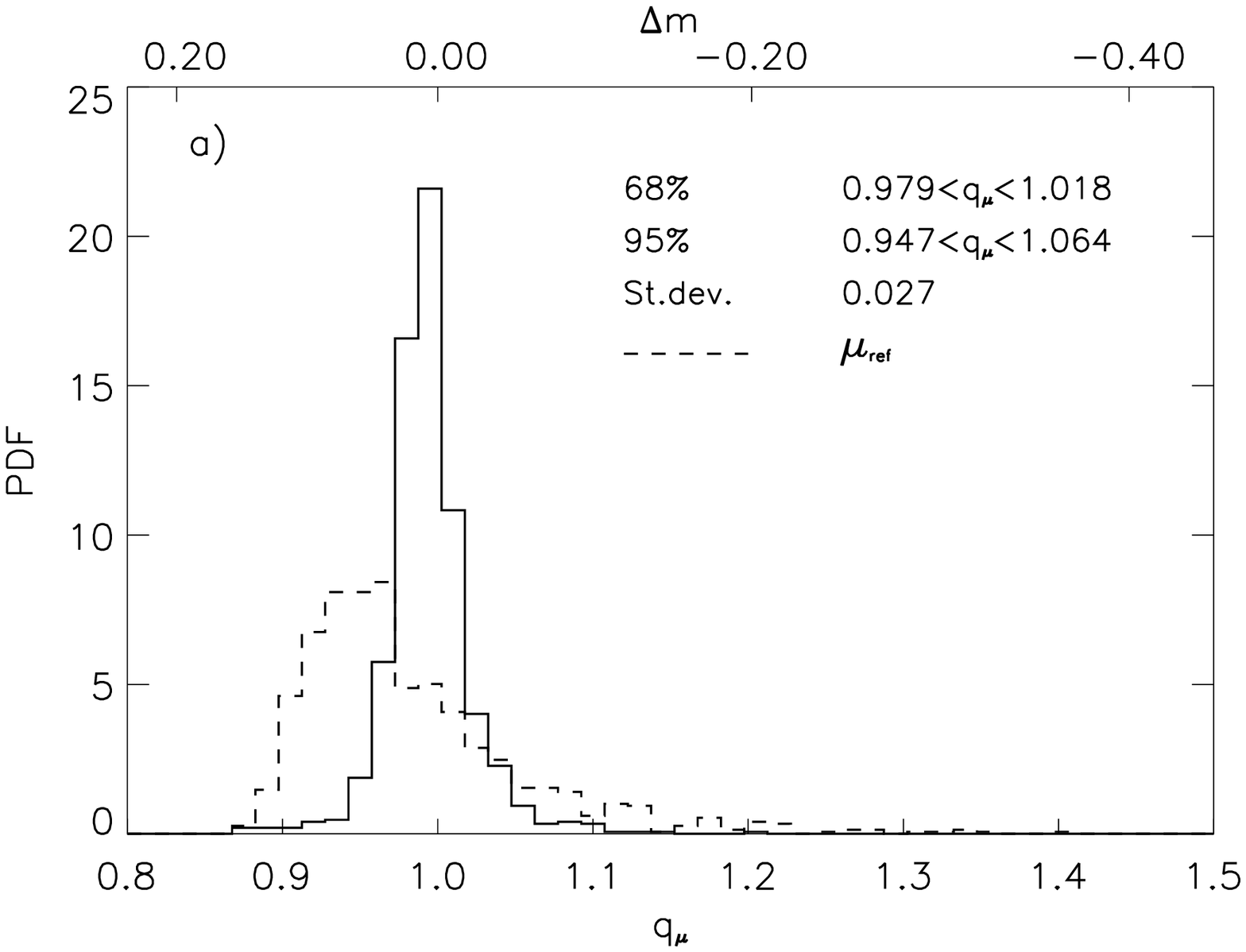}}
\resizebox{0.45\textwidth}{!}{\includegraphics{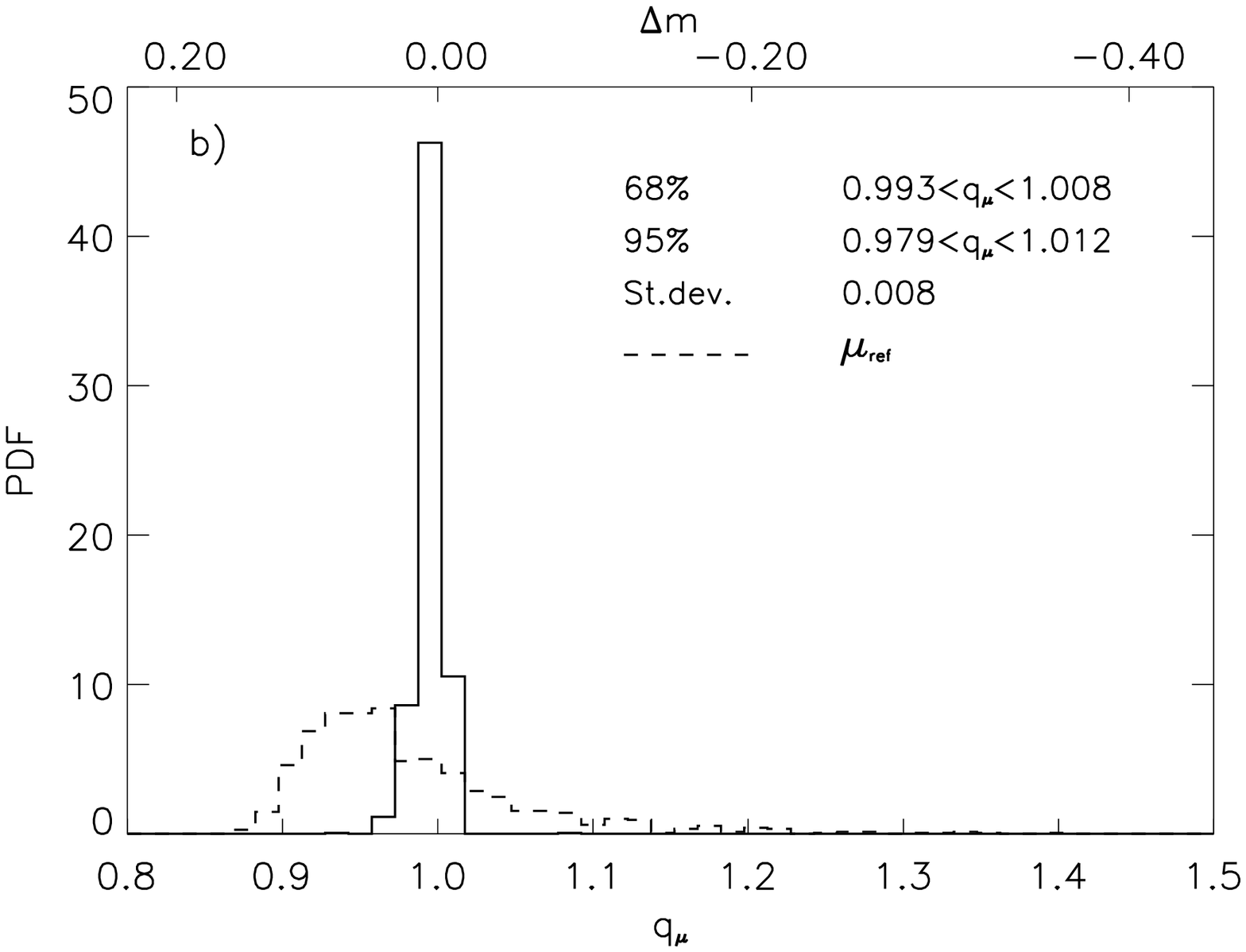}}
\resizebox{0.45\textwidth}{!}{\includegraphics{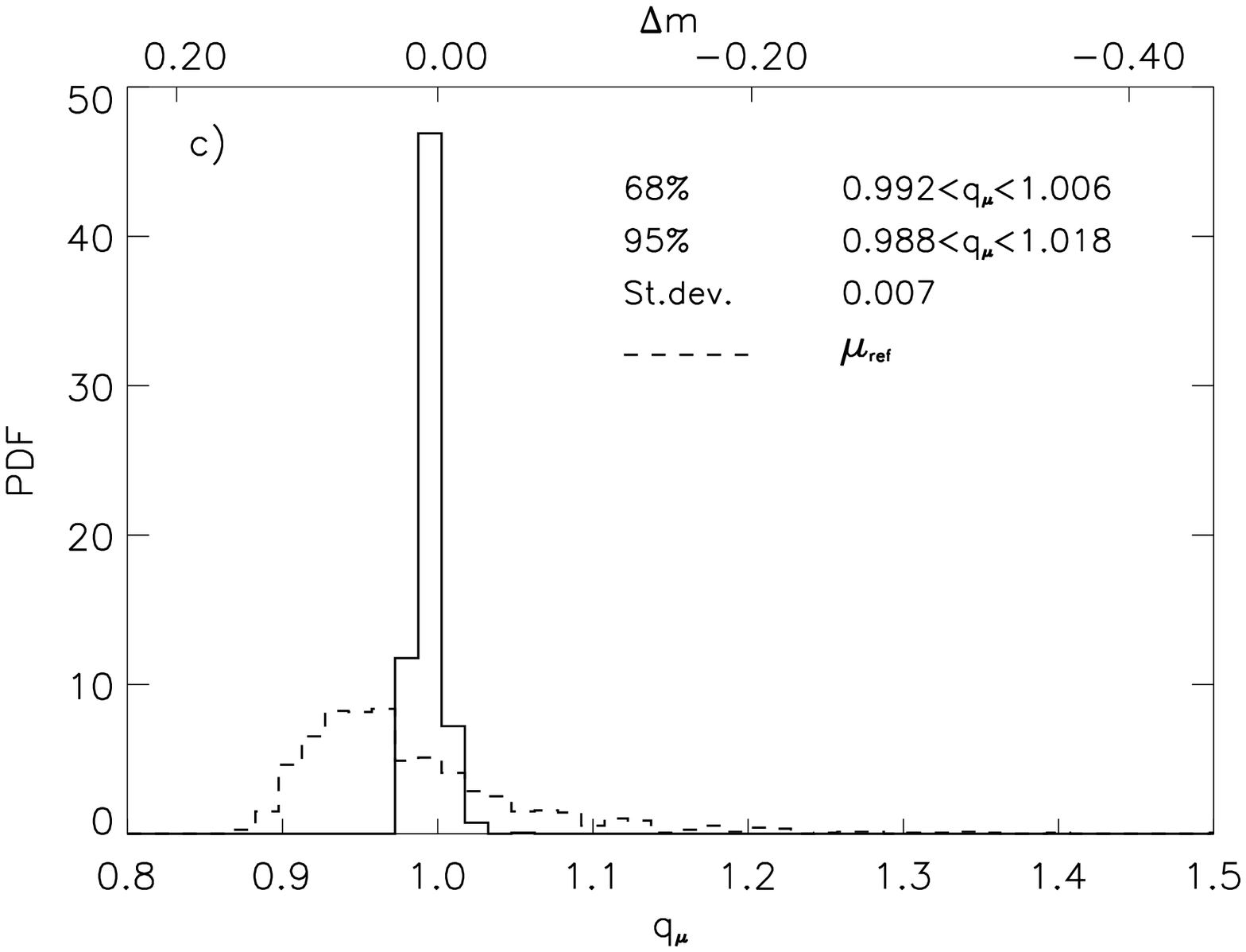}}
\resizebox{0.45\textwidth}{!}{\includegraphics{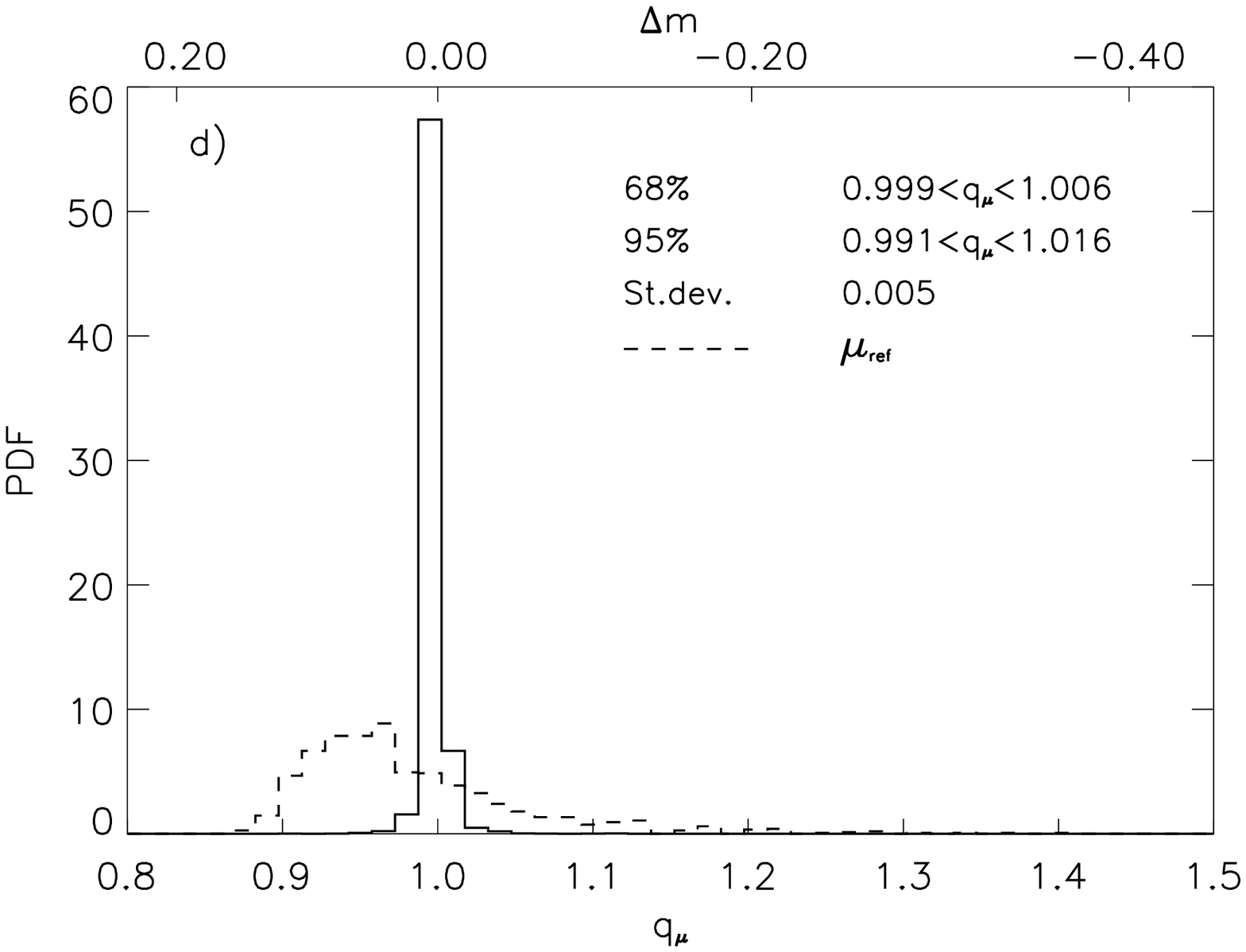}}
\resizebox{0.45\textwidth}{!}{\includegraphics{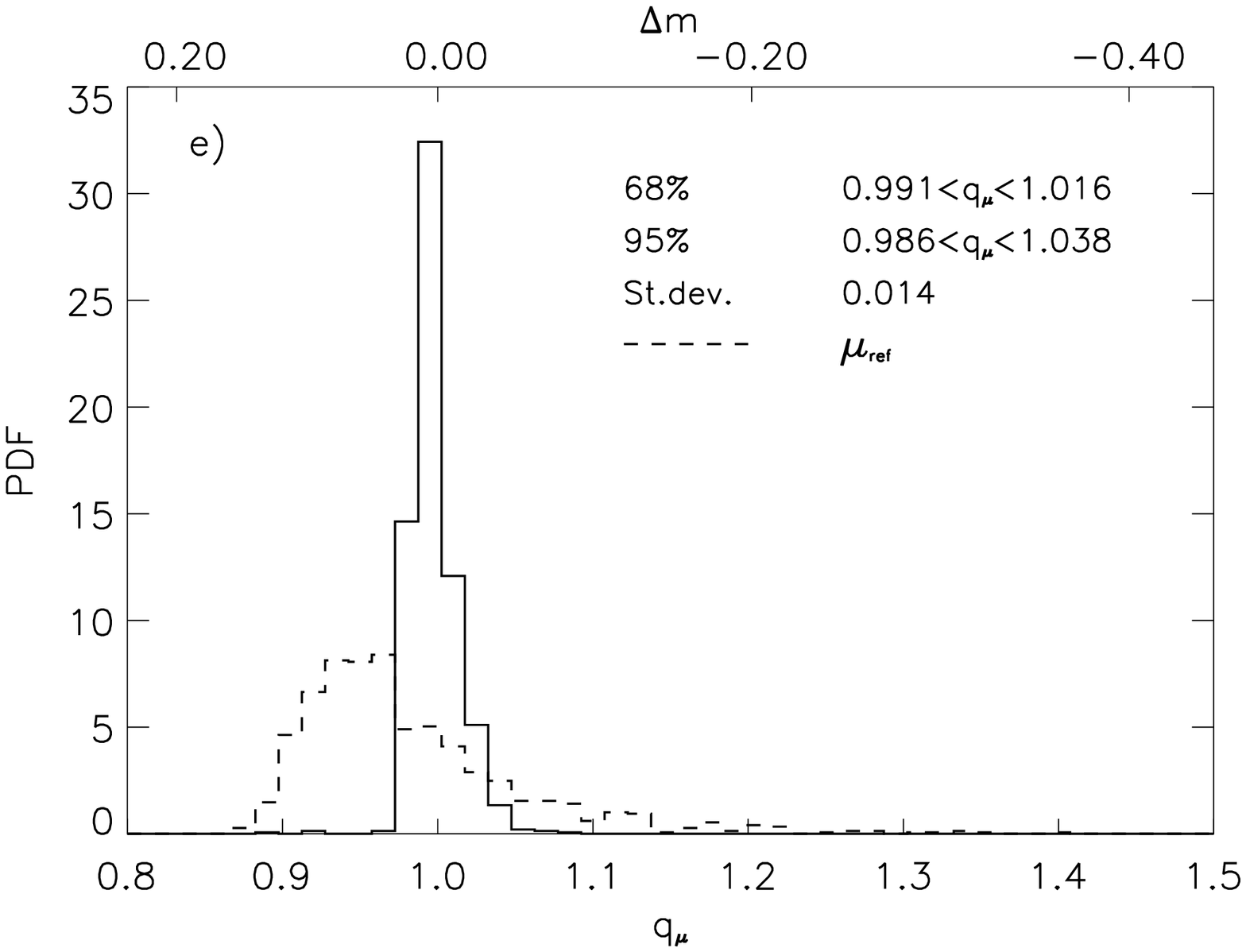}}
\resizebox{0.45\textwidth}{!}{\includegraphics{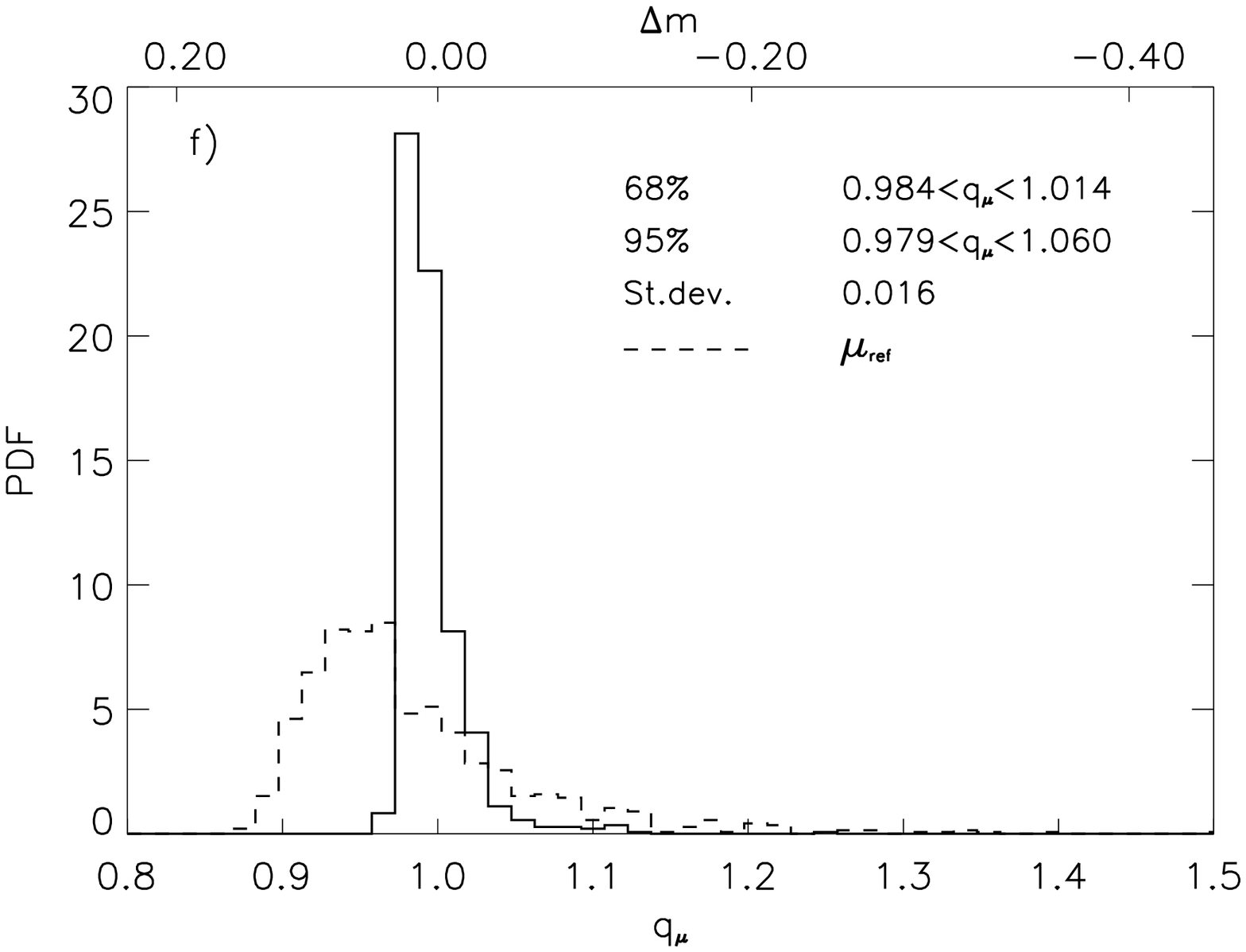}}
\caption{\label{fig:pdffigs} PDFs of $q_{\mu}$ where the
  following uncertainties have been studied: a) The scatter in the velocity dispersion
  of the F--J and T--F relations, b)
  Offset in central value of $\sigma_*$ of $+10$ km$\ \!$s$^{-1}$, c) Offset in
  central value of $\sigma_*$ of $-10$ km$\ \!$s$^{-1}$, d) Lens redshifts
  distributed around their original values, e) Assuming SIS halos
  instead of NFW and f) Assuming a magnitude limit of $I=23$ mag. The
  PDF before correction (dashed line) is also plotted for reference.}
\end{center}
\end{figure}

\begin{figure}
\begin{center}
\resizebox{0.4\textwidth}{!}{\includegraphics{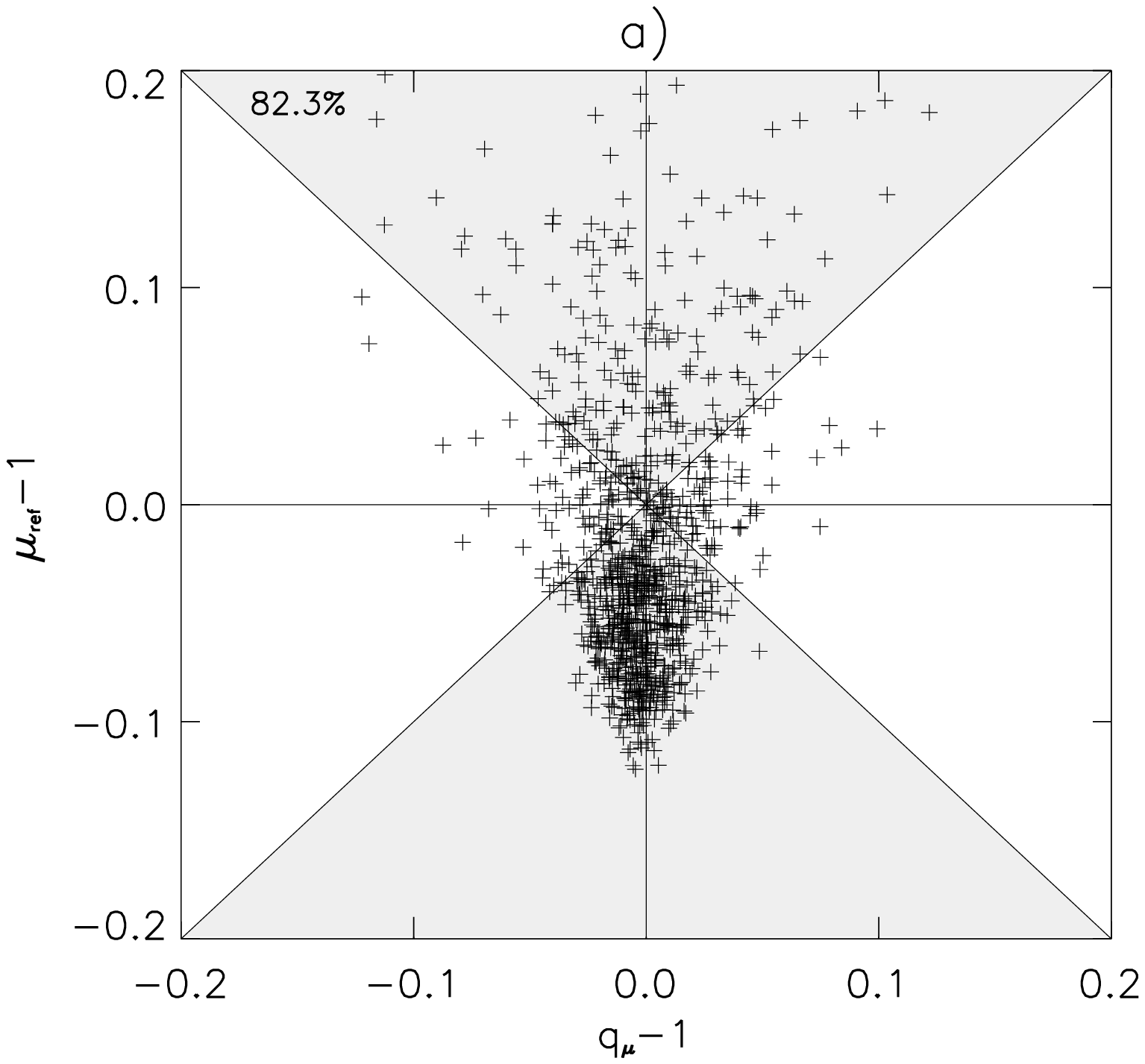}}
\resizebox{0.4\textwidth}{!}{\includegraphics{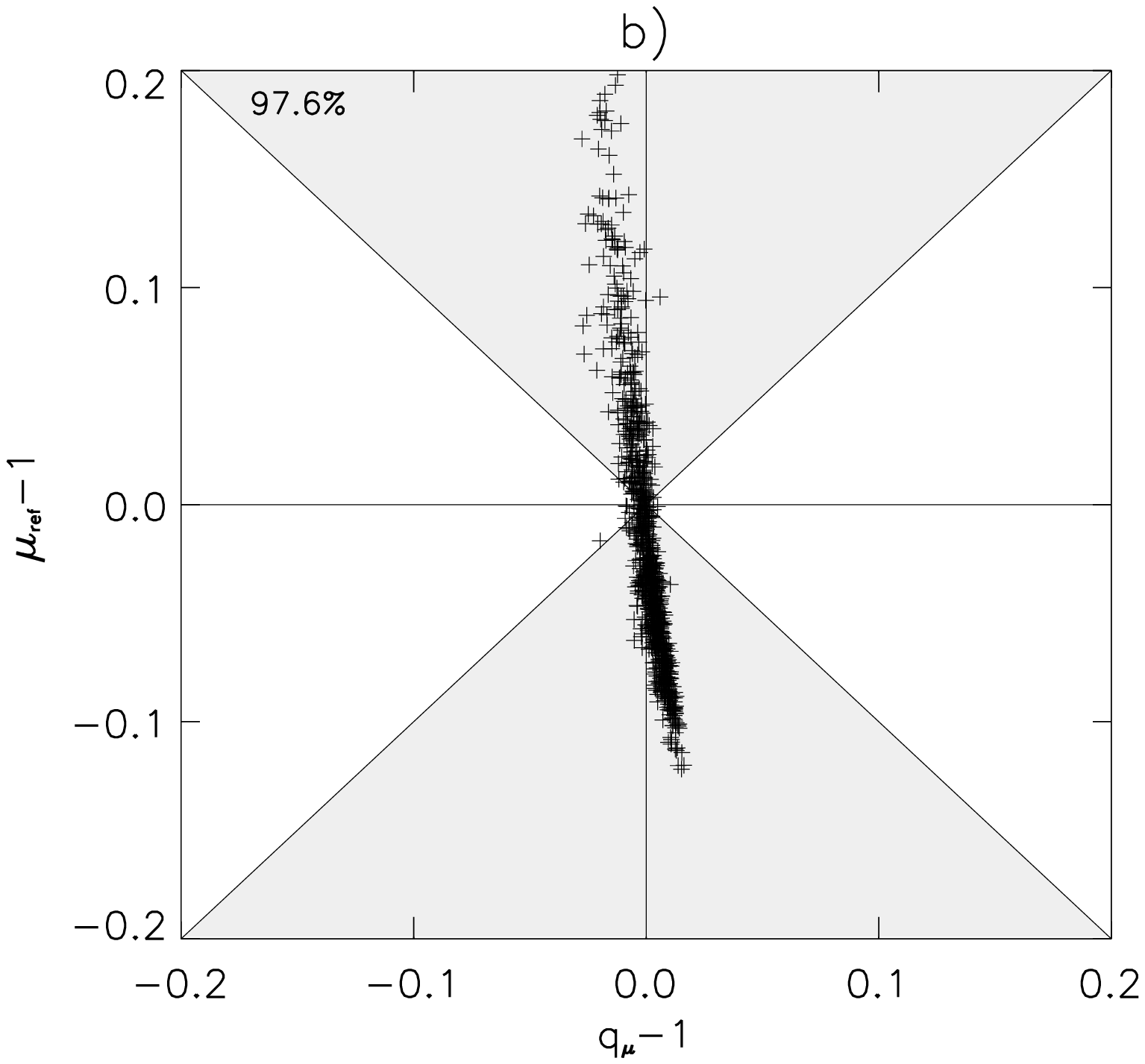}}
\resizebox{0.4\textwidth}{!}{\includegraphics{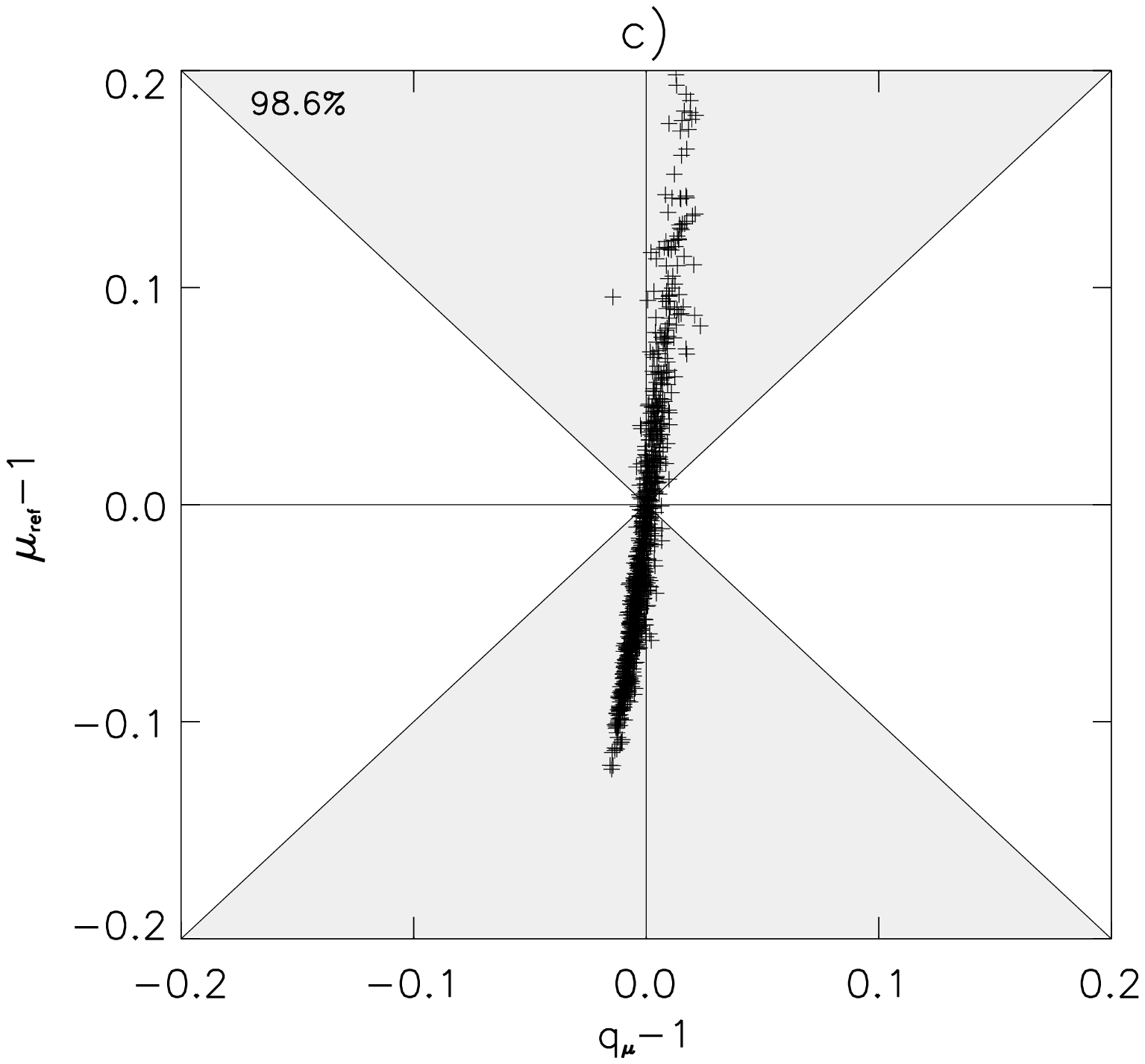}}
\resizebox{0.4\textwidth}{!}{\includegraphics{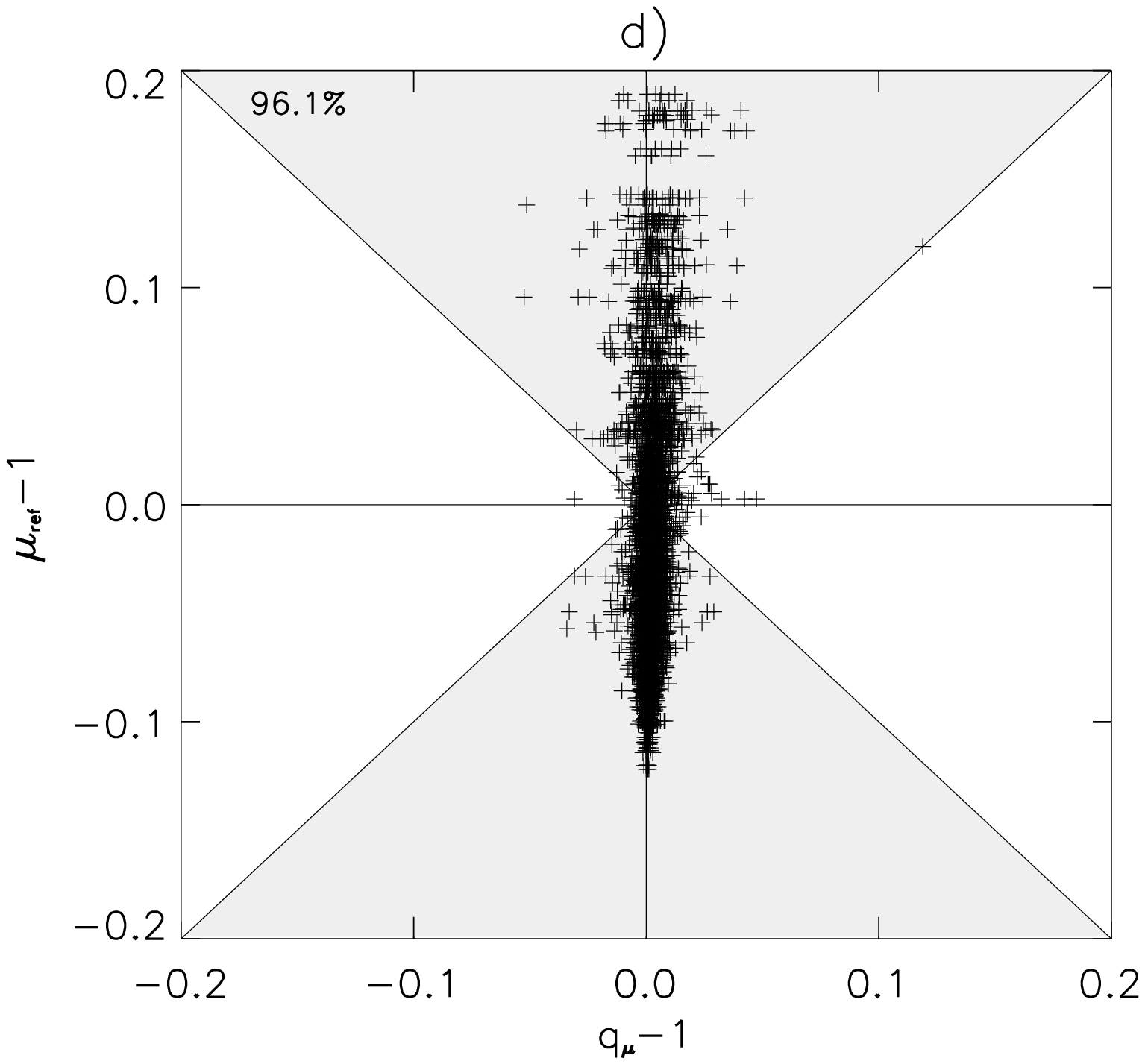}}
\resizebox{0.4\textwidth}{!}{\includegraphics{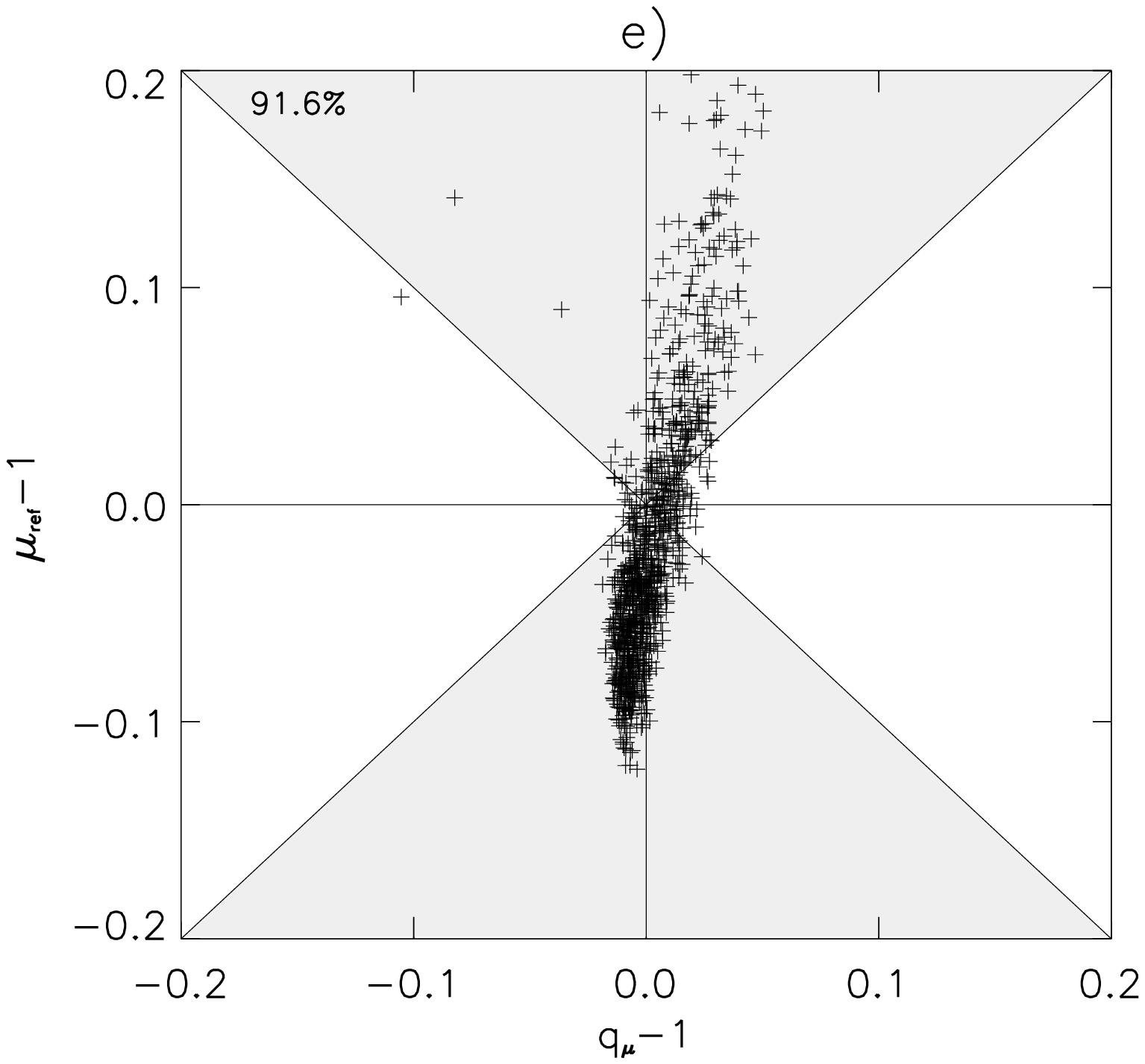}}
\resizebox{0.4\textwidth}{!}{\includegraphics{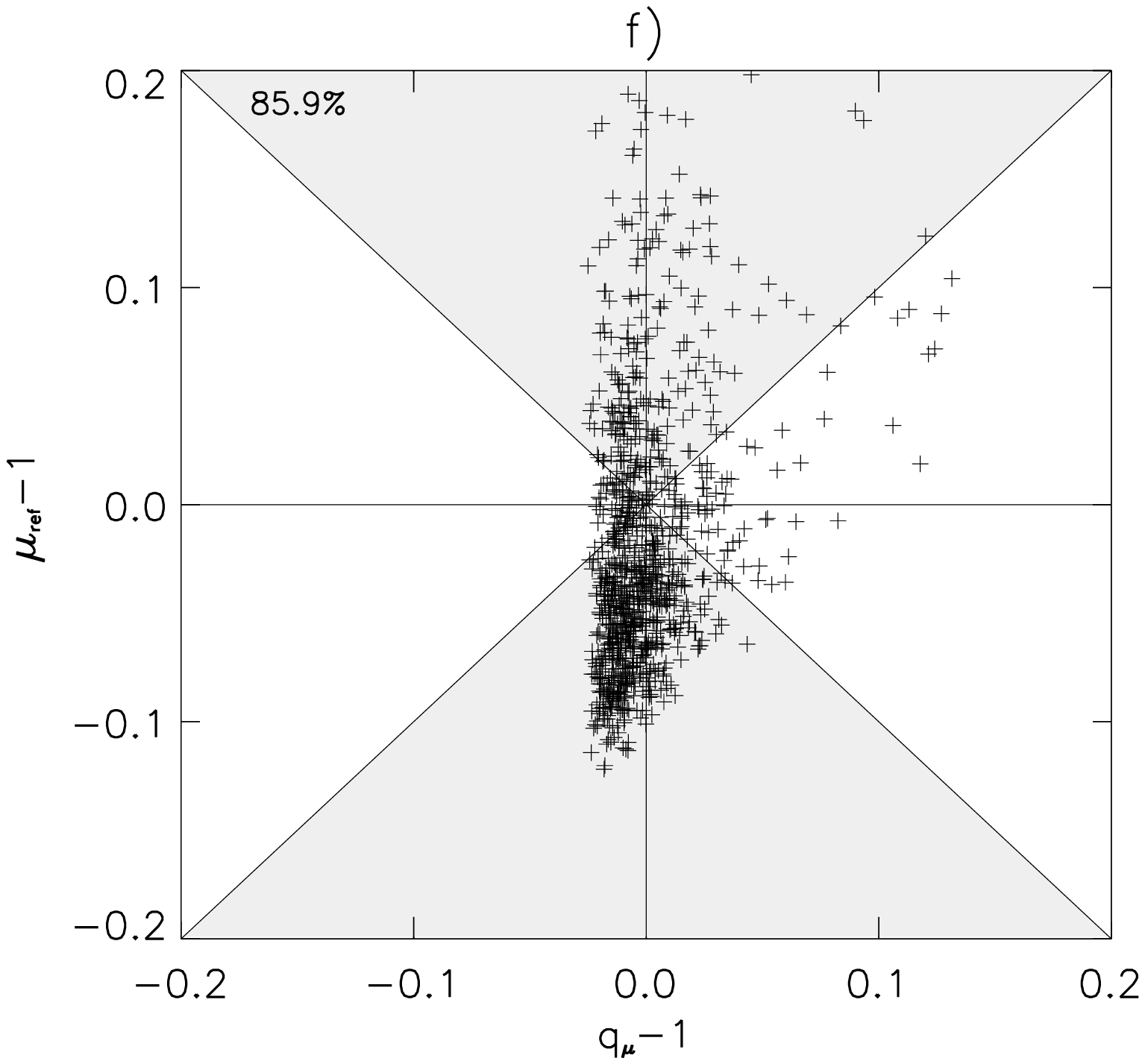}}
\caption{\label{fig:spreadfigs} Spread in $\mu_{\rm ref}-1$ vs spread in
  $q_{\mu}-1$ where the following uncertainties have been studied: a)
  The scatter in the velocity dispersion
  of the F--J and T--F relations, 
  b) Offset in central value of $\sigma_*$ of $+10$
  km$\ \!$s$^{-1}$, c) Offset in central value of $\sigma_*$ of $-10$ km$\ \!$s$^{-1}$, d)
  Lens redshifts distributed around their original values, e) Assuming
  SIS halos instead of NFW and f) Assuming a magnitude limit of $I=23$
  mag. The gray shaded areas show where the value after correction is
  closer to reality than before correction. The figures in the upper
  left corners give the percentage of events residing in the gray
  areas, where the correction improves the resulting accuracy of the
  derived distance.}
\end{center}
\end{figure}

In panel a) in Figure~\ref{fig:pdffigs}, we compare the distribution
of the corrected value $q_\mu$ due to the scatter in the F--J and T--F
relations with the distribution of
magnifications in the reference model (dashed line). Note that if we
knew all velocity dispersions and circular velocities exactly, $q_\mu$
would be represented by a 
$\delta$-function at $q_{\mu}=1$. We see that the intrinsic scatter in the
 velocity dispersion and circular velocity
causes a dispersion of $q_\mu$ of approximately $3\,\%$, a factor of
$\sim 2.6$
less than the original dispersion. Panel a) in
Figure~\ref{fig:spreadfigs} shows $\mu_{\rm ref}-1$ vs $q_{\mu}-1$ for
each individual source. For $82\,\%$ of the sources, the corrected
luminosity will be better than the uncorrected one. 

Besides the intrinsic scatter in the F--J and T--F relations, there is
also a possible \emph{systematic} uncertainty in $\sigma_*$. To
estimate this, we use SDSS data
in \citet{bernardi03} where 
photometric and spectroscopic parameters for a
sample of $\sim 9000$ early-type galaxies in the redshift range
$0.01<z<0.3$ are given, including K-corrections and accurately
measured velocity dispersions. We fit a straight line $M_B=b\log\sigma
+a$ and estimate the error in $\sigma_*$ for a given $M_B^*$ by
propagating the errors in the parameters $a$ and $b$.
For $M_B^*=-21.04$, we obtain $\sigma_*=218\pm 7$ km$\ \!$s$^{-1}$. In panel b)
and c) in Figure~\ref{fig:pdffigs}, we investigate the effect of a
systematic shift of $\pm 10$ km$\ \!$s$^{-1}$ in $\sigma_*$. The dispersion in
$q_\mu$ due to such a shift is quite small or at the order of $1\,\%$.
Panels b) and c) in Figure~\ref{fig:spreadfigs} shows $\mu_{\rm ref}-1$
vs $q_{\mu}-1$ for each individual source. The corrected value will be
better than the uncorrected for $>95\,\%$ of the sources.

A further source that may increase the scatter in the magnification is
the possible misclassification of galaxy morphology. An elliptical galaxy
wrongly classified as a spiral, leads to an underestimation of the underlying
mass, and vice versa if a spiral is misclassified as an elliptical. To
investigate the possible effect of this, we first use a set of simulated galaxies
with known morphological types (i.e.~an exponential radial profile for 
spirals and a de Vaucouleurs profile for ellipticals) and measure how many are 
correctly recovered in an observational setup resembling the GOODS. 
To classify galaxies, we use the GALFIT software
\citep{peng02}, which measures the slope of the radial profile and 
therefore allows a discrimination between ellipticals and spirals.
At a S/N=10 detection limit (m$\sim$25), we find that $\sim$25\,\% of the 
ellipticals and $\sim$8\,\% of the spirals are misclassified. The fraction
of misclassified galaxies quickly drops to $\sim$1\,\% at a magnitude
$\sim$2 mag brighter than the detection limit. We then use these results 
to estimate the effect of misclassification on the derived magnifications. 
We find that the increase in the scatter in the magnification
is $\sim$0.5\,\%~due to this effect. Therefore, misidentification of 
galaxy morphology should only affect the results marginally. However,
for ground-based surveys with low resolution, the effect may be larger.

\subsection{Redshift and Position Uncertainties}\label{subs:reduncert}
Uncertainties in the redshifts of the lenses will alter the results
both through the uncertain distances between different lens-planes and
by introducing an uncertainty in their absolute magnitudes used in the
F--J and T--F relations.

In an ideal observational situation, all redshifts are determined
spectroscopically. In many real situations, however, only photometric
redshift are available due to, e.g., the faintness of the galaxies and
the large number of sources. To investigate the effects of photometric
redshift uncertainties, we add a random offset to the redshift of each
object. The size of the offset depends on redshift, apparent
magnitude and spectral type of the object and is drawn from the simulated 
error distribution discussed above.

In panel d) in Figure~\ref{fig:pdffigs}, we show the distribution of
the corrected value $q_\mu$ due to a random offset to the redshift of
each lensing object. The induced error in the estimated magnification
is less than $1\,\%$. The corresponding panel in Figure~\ref{fig:spreadfigs}
shows $\mu_{\rm ref}-1$ vs $q_{\mu}-1$ for each individual
source. In this case, $96\,\%$ of the sources have corrected values
which are better than when uncorrected.

Besides the Gaussian-like distribution of the photometric redshift errors
investigated above, there is also a possibility that a fraction of
the objects get "catastrophic redshifts" with large errors, so called 
outliers. E.g., by comparing with $\sim$1400 spectroscopic redshifts in the 
CDF-S and HDF-N, we find that the GOODS photometric redshifts have 
about 3\,\% outliers with $\Delta_z>0.3$. The redshift probability 
distribution, for a majority of these objects are characterized by a 
primary ($\sim$Gaussian) peak combined with a less pronounced secondary peak. 
Outliers are foremost objects assigned
the redshift of the primary peak, but where the true redshift is that of
the secondary peak. To estimate the effect of outliers, we use the galaxies
in the GOODS and simulate the case where we distribute the photometric 
redshifts over the full probability distribution, including the secondary 
peak. This will allow $\sim$3\,\% outliers. We compare
this with the case where we only include redshifts in the primary peak
(i.e.~objects with $\Delta_z<0.3$). We find that the increase in lensing 
dispersion due to the population of outliers is less than 1\,\%. One reason
for the small effect is that outliers are mainly faint and very blue
objects, which therefore should have relatively small masses. 

Any error in the exact positions of the lensing galaxies can also
affect the resulting magnification. Apart from the observational
error, such an effect can be due to a misalignment between the
luminous and the dark matter in a given galaxy. We have investigated
the effect of a Gaussian random shift with $\sigma_{\rm pos}=0.5$
arcseconds of all lensing galaxies along the l-o-s. Even such a large
shift of {\em all} galaxy positions results in a distribution of
corrected values $q_\mu$ with a dispersion of less than
$0.5\,\%$. 

\subsection{Choice of Halo Profile}\label{subs:halochoice}
The choice of halo model is only important in those lens-planes where
the light-ray passes through a halo. If passing outside, the halo will
act as a point mass and when $m_{\rm tot}^{\rm halo}=m_{200}^{\rm
SIS}=m_{200}^{\rm NFW}$ the two different halo models give exactly the
same results.

We have performed simulations where all halos were of SIS 
instead of NFW type. The effect of different halo profiles 
are also present in the realistic and pessimistic case simulations
below.

Panel e) in Figure~\ref{fig:pdffigs} shows the PDF of $q_\mu$ when
assuming SIS halos instead of NFW as in the reference model. The
dispersion is less than $1.5\,\%$. In $>90\,\%$ of the cases, the
corrected value is better than the uncorrected, see panel e) in
Figure~\ref{fig:spreadfigs}.

We have also investigated how important the assumption on the
truncation radius is for the resulting magnification distribution by
running tests with $0.75\times r_{200}\leq r_{\rm t} \leq 1.25\times
r_{200}$ for the SIS model. The uncertainty in the resulting
magnification gives a distribution of corrected values $q_\mu$ with a
dispersion of $\sim 0.5-1\,\%$. Since the NFW profile falls of as
$\rho_{\rm NFW}\propto r^{-3}$ at large radii as compared to
$\rho_{\rm SIS}\propto r^{-2}$ for SIS halos, this should be
considered a very conservative limit on the effect of changing the
truncation radius.

\subsection{Magnitude Limits}\label{subs:maglimsims}
Our reference model uses a constant comoving mass density of galaxies,
implying a constant smoothness parameter, $\eta$. In a real scenario
with an observational magnitude limit, an increasing fraction of
galaxies drop out at higher redshift. The faint high redshift galaxies
will not be seen and hence not included as lenses in the magnification
calculation but instead added as homogeneously distributed
matter. Therefore, when deriving the smoothness parameter from
observations, $\eta(z)$ increases with redshift even if the
'underlying' smoothness parameter is constant. For each simulation
with a finite magnitude limit, a new $\eta$-function is computed using
the method described in \S\ref{subs:smooth}.

For $I=27$ mag the distribution of $q_{\mu}$ is very narrow and for
$I=29$ mag, it is in principle a $\delta$-function.  Panel f) in
Figure~\ref{fig:pdffigs} shows the PDF of $q_\mu$ for $I=23$. The
dispersion is $\sim 2\,\%$. In $86\,\%$ of the cases, the
corrected value is better than the uncorrected, see panel f) in
Figure~\ref{fig:spreadfigs}.
 
\begin{figure}
\begin{center}
\resizebox{0.45\textwidth}{!}{\includegraphics{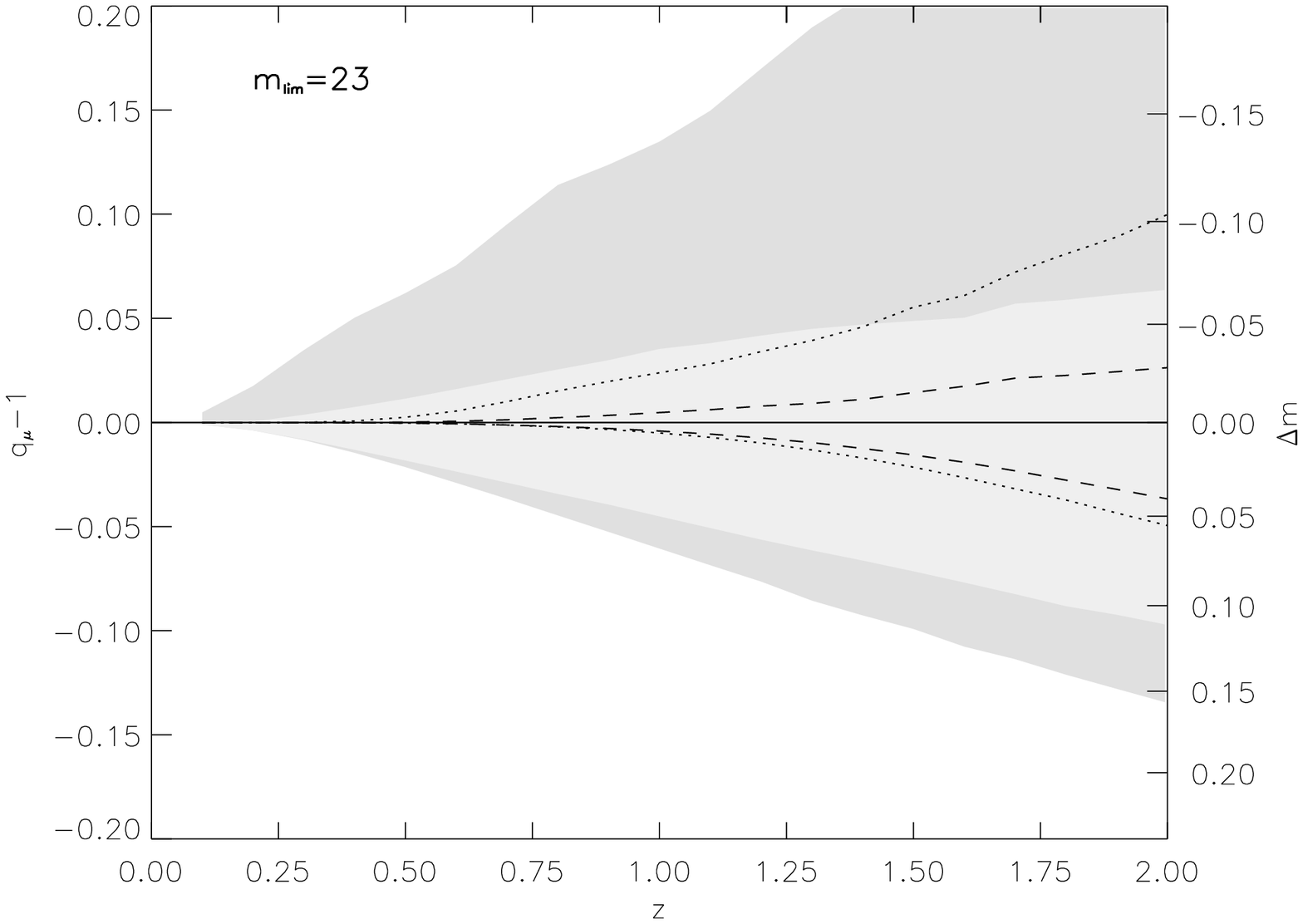}}
\resizebox{0.45\textwidth}{!}{\includegraphics{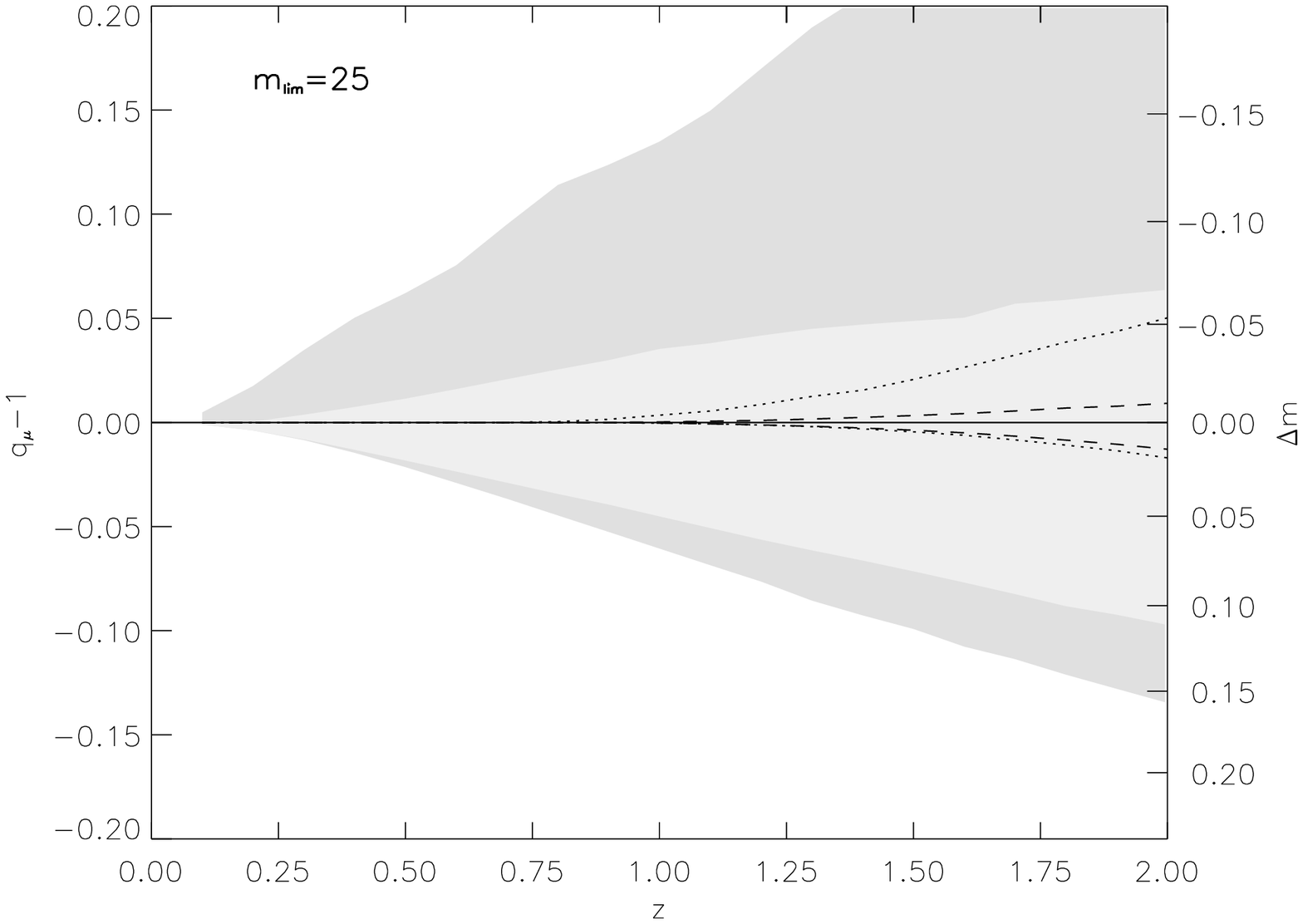}}
\caption{\label{fig:incompl2325}  $q_{\mu}-1$ as a function of source redshift
  for a models with magnitude limits $I=23$ and $I=25$, using 1700
  lines-of-sight for each redshift. The gray shaded areas show the
  68\,\% and 95\,\% confidence levels of $\mu_{\rm ref}-1$, the distance
  from the correct value 1 before correction. The dashed and dotted
  lines show the same confidence levels after correction clearly
  indicating the improvement.}
\end{center}
\end{figure}

In Figure~\ref{fig:incompl2325}, we compare $q_{\mu}-1$ as a function of
source redshift for models with magnitude limits $I=23$ and $I=25$
with $\mu_{\rm ref}-1$. Even for source redshifts as high as $z\sim
2$, a magnitude limit of $I=23$ does not significantly impair our
results.

Photometric errors in the apparent magnitudes translates to an
increased scatter in the absolute magnitudes and therefore also in the
derived velocity dispersions and masses. At the faintest magnitude 
limits considered here, S/N=10, typical errors are $\sim 0.1$~mag.
For ellipticals, this corresponds to an increased dispersion of
$\Delta{\rm log_{10}}\sigma \sim 0.01$ (using Eq. 10). This is 
significantly less than the intrinsic scatter in the F--J relation 
which is rms(log$_{10})\sigma \sim 0.08$~(at $\sim M_B^*$, Eq. 12). For the
T--F relation the increased scatter due to photometric errors is
$\Delta{\rm log_{10}}V_{\rm max}\sim 0.013$ (using Eq. 13), again
significantly less than the intrinsic scatter
 rms(log$_{10}V_{\rm max}) \sim 0.06$.
So even for the faintest galaxies considered, the errors in apparent 
magnitude should not affect results more than marginally.
  
\subsection{Realistic and Pessimistic Scenarios}\label{subs:wcrc}
We have studied the uncertainty in the lensing correction in a
realistic scenario where a reasonable error budget is assumed. In
this case we have assumed 50\,\% NFW, 50\,\% SIS, no central value shift
of the velocity dispersion normalization but a dispersion around this
value and a magnitude limit of $I=25$. Lens redshifts were assumed to
be distributed around their reference values. As the correct model we have
used the reference NFW model as above.

A pessimistic scenario for space based surveys has also been
studied where we have 
maximized the uncertainties in relating the luminous and the dark
matter. However, one could easily imagine worse cases in a
ground-based experiment. 
For our scenario the erroneous assumptions were: SIS halos, central value of
velocity dispersion normalization shifted +10 km$\ \!$s$^{-1}$ and distributed
around this value, a magnitude limit of $I=25$, lens redshifts assumed
to be distributed around their reference model values, an offset in lens
positions as described in \S\ref{subs:reduncert} and finally a
truncation radius of $1.25\times r_{200}$. We consider this being a
pessimistic but not completely unrealistic scenario.

\begin{figure}
\begin{center}
\resizebox{0.45\textwidth}{!}{\includegraphics
  {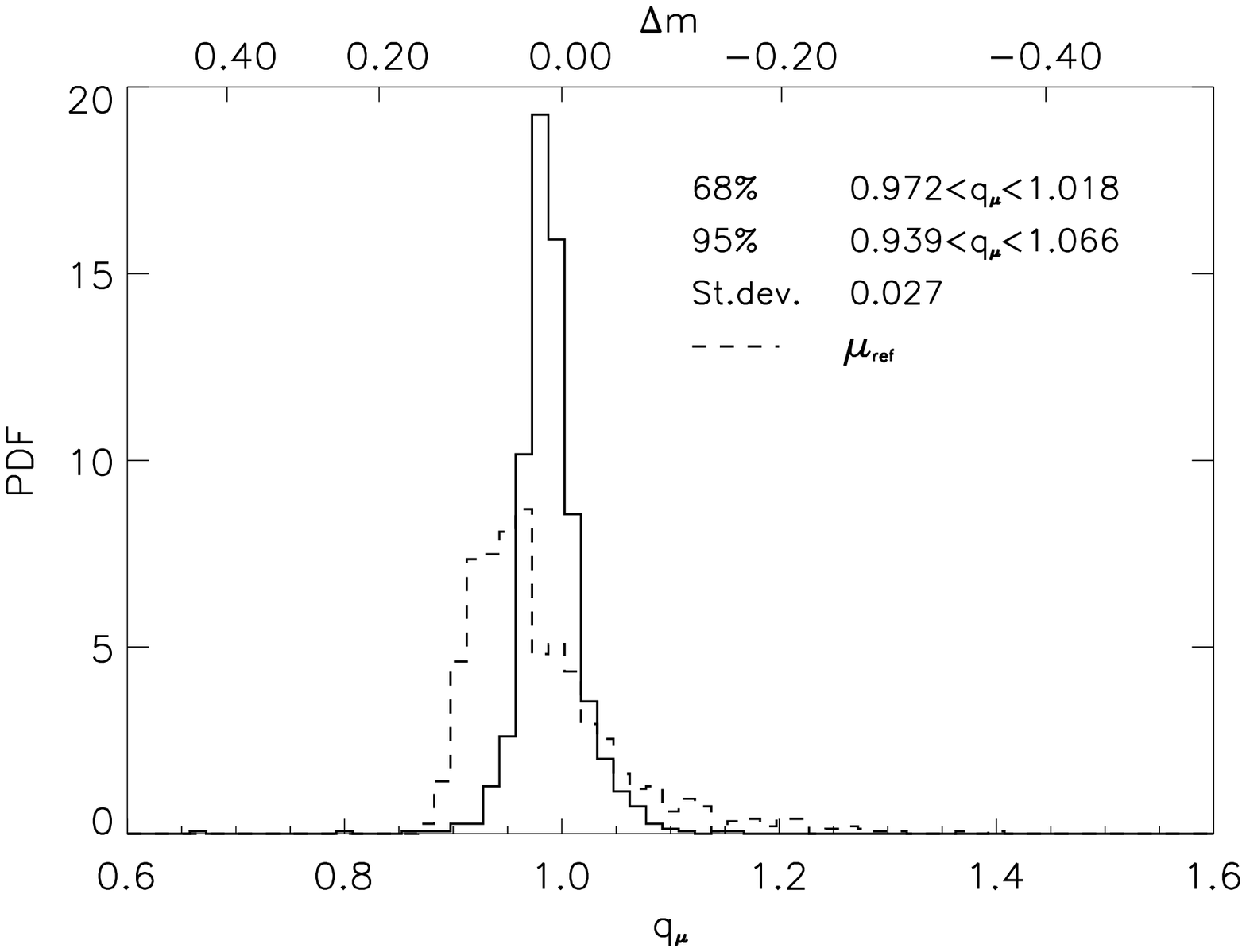}} 
\resizebox{0.45\textwidth}{!}{\includegraphics
  {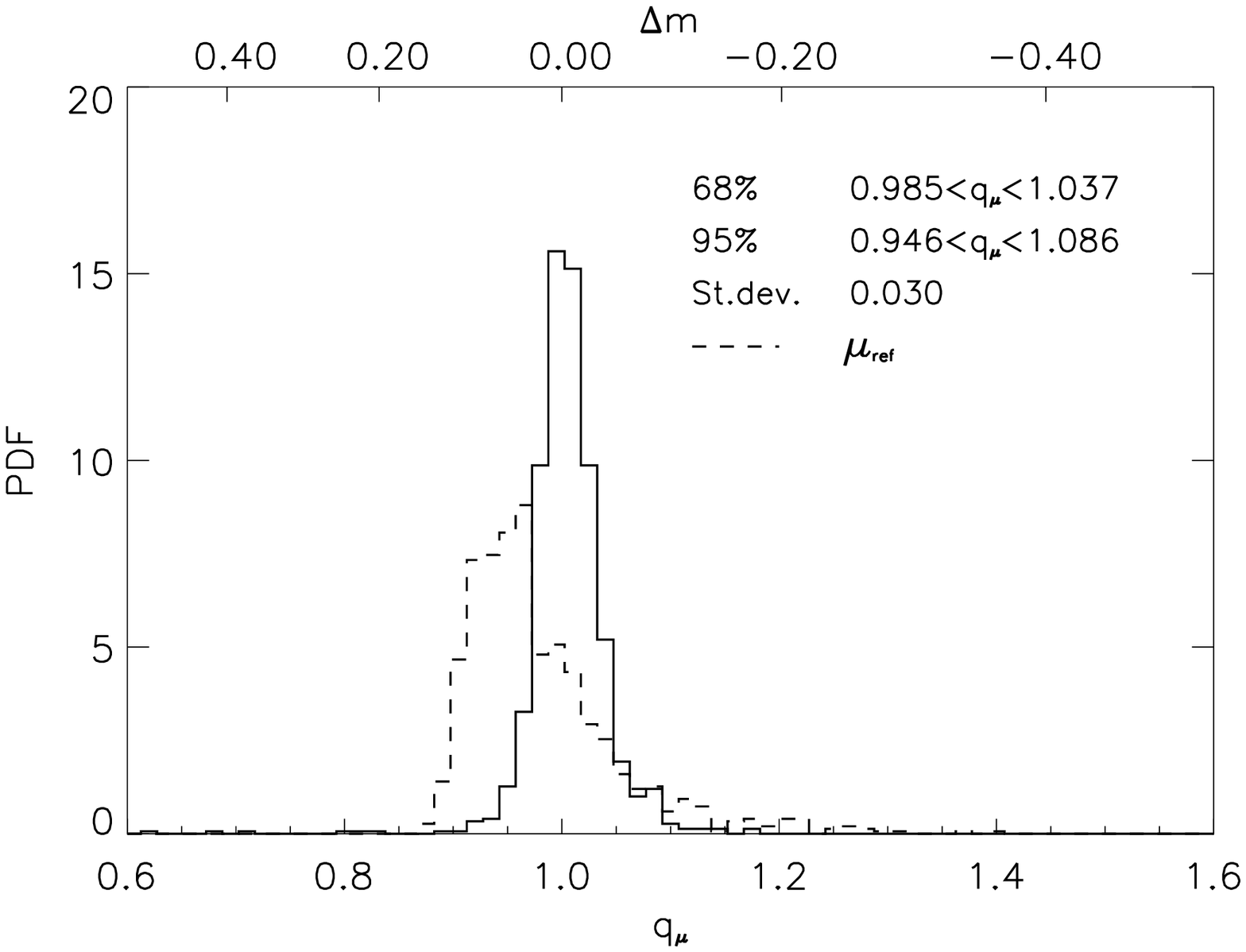}}
\resizebox{0.45\textwidth}{!}{\includegraphics
  {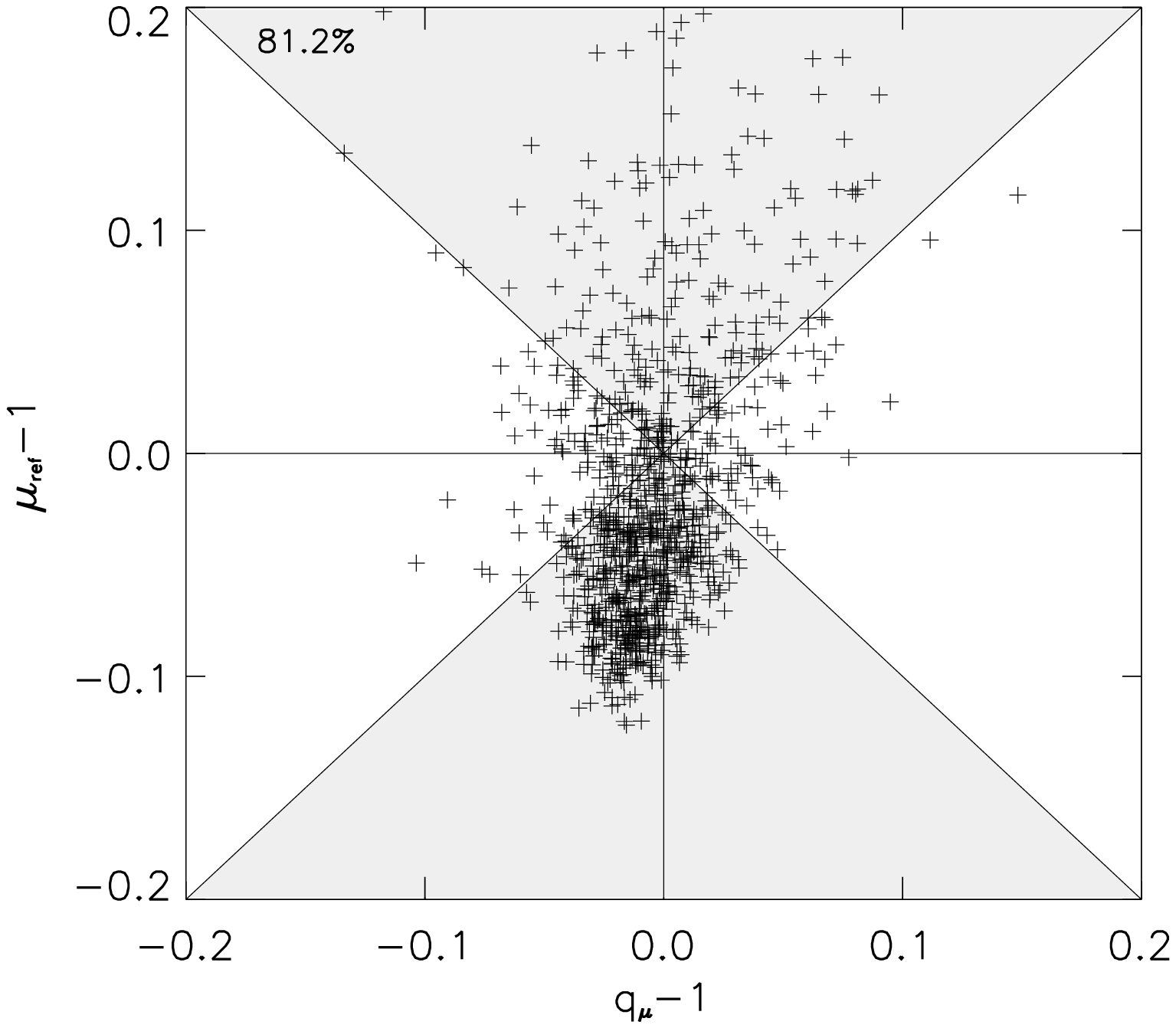}}
\resizebox{0.45\textwidth}{!}{\includegraphics
  {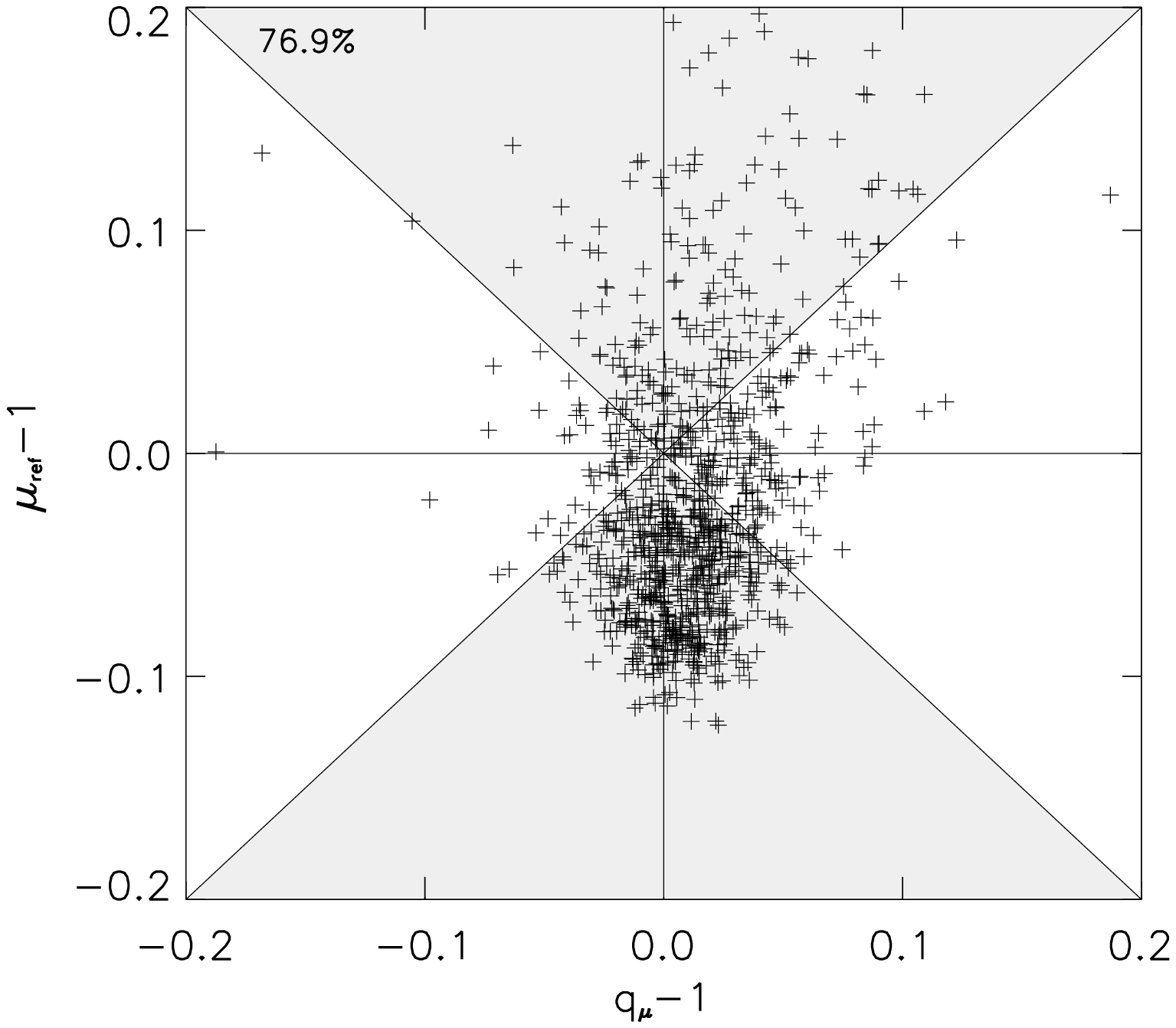}} 
\resizebox{0.45\textwidth}{!}{\includegraphics
  {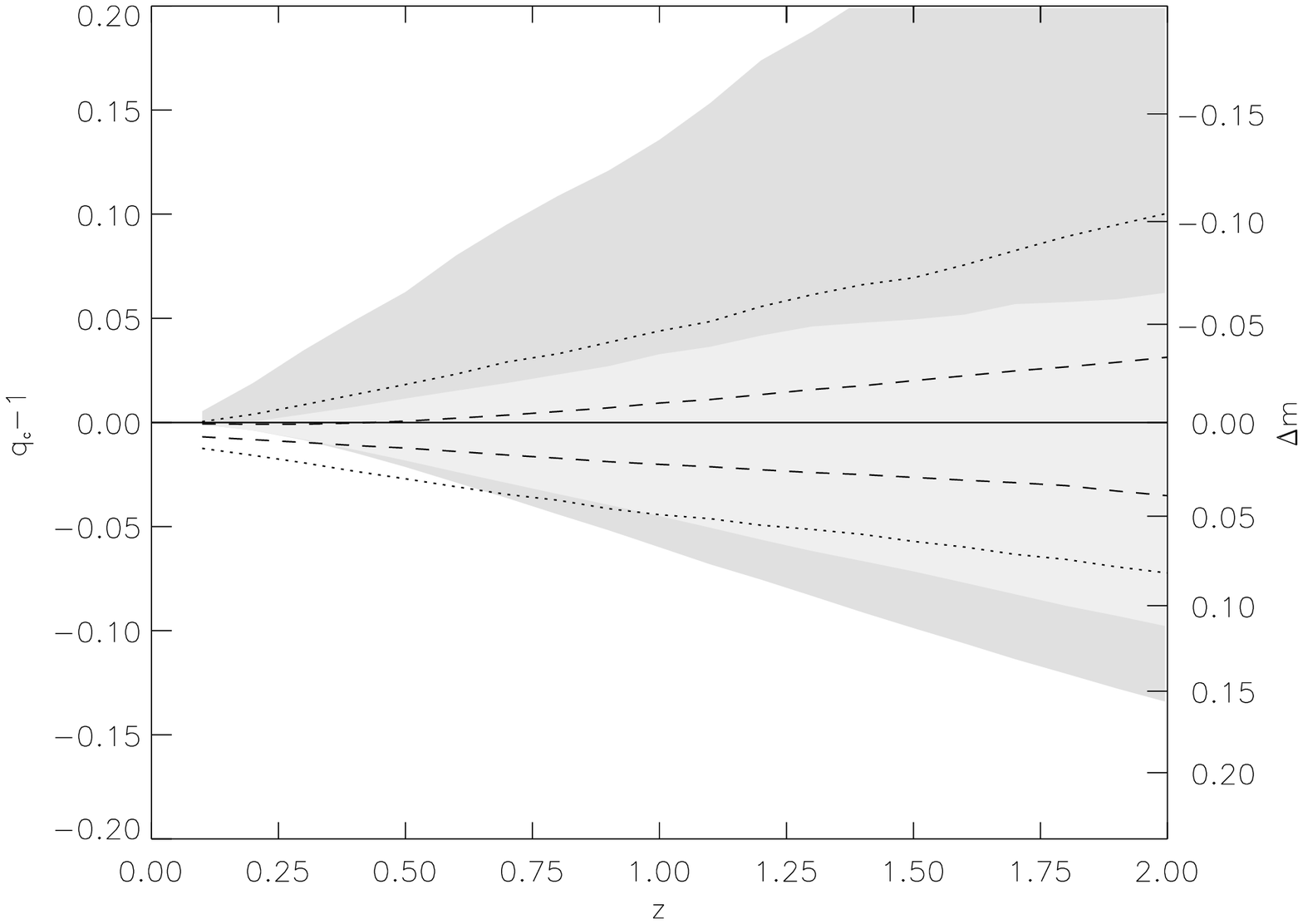}}
\resizebox{0.45\textwidth}{!}{\includegraphics
  {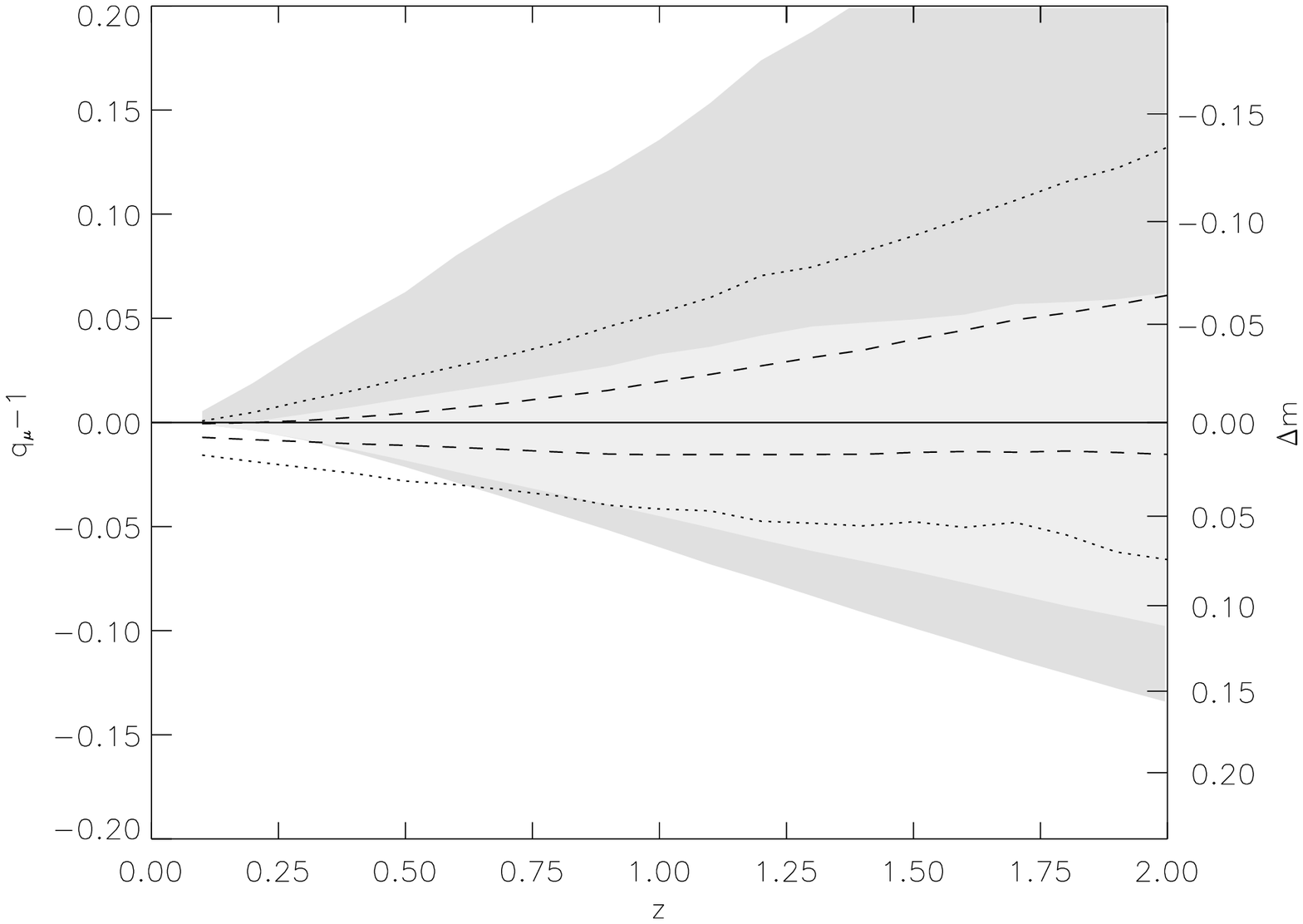}}
\caption{\label{fig:wcrc} Left column
  shows a realistic scenario with 50\,\% NFW, 50\,\% SIS, values of
  velocity dispersion distributed around their central value, lens
  redshift values distributed around their reference model values and
  finally using a magnitude limit of $I=25$ mag. 
  The right column shows a pessimistic scenario with
  with 100\,\% SIS, values of velocity dispersion
  normalization distributed around a +10 km$\ \!$s$^{-1}$ shifted central value, a
  small random offset $(\sigma_{\rm pos}=0.5'')$ in lens positions,
  truncation radius of $1.25\times r_{200}$
  and the rest as in the realistic case.
  The upper row gives the PDFs and confidence levels
  for sources at redshift $z=1.5$, middle row shows the spread in
  $\mu_{\rm ref}-1$ vs the spread in $q_{\mu}-1$.  The bottom row shows
  confidence levels of the two scenarios vs source
  redshift.}
\end{center}
\end{figure}

The left panels in Figure~\ref{fig:wcrc} show results for the realistic
scenario, the right panels for the pessimistic scenario. For the realistic
case, $q_{\mu}$ has a dispersion of $\sim 3\,\%$ and
$(q_{\mu}-1)<(\mu_{\rm ref}-1)$ for $\sim 80\,\%$ of the sources. In
the pessimistic case, the corresponding numbers are $\sim 3\,\%$ and $\sim
77\,\%$, i.e.~our ability to correct for lensing is more or less
unimpaired when going from a realistic to a pessimistic scenario. The
bottom row shows $q_{\mu}-1$ as a function of source redshift for the
two scenarios. The confidence levels of the realistic case vs $z$ can
be well fitted with straight lines and these are found in Table
\ref{tab:lines} expressed in magnitudes.

\section{SUMMARY AND DISCUSSION}\label{sec:summary}
We have investigated the accuracy to which lensing magnification can
be estimated on individual lines of sight using the observed properties of
the foreground galaxies of each source. 

The result depends on the uncertainties in translating observed galaxy
luminosities to the (invisible) matter distribution in the lensing
galaxies. We have shown that none of the studied uncertainties neither
individually nor combined will render the corrected distribution of
magnifications wider than the dispersion from lensing. Even for a
pessimistic scenario, the dispersion due to lensing for a standard
candle source at $z=1.5$ can be reduced by a factor $\gtrsim 2$,
comparable to the result for a realistic scenario. The reason our
pessimistic case result is not significantly worse than for the
realistic case, is that the uncertainties are dominated by the scatter
in the F--J and T--F relations for both scenarios\footnote{We were
able to increase the width in the $q_{\mu}$ distribution for the
pessimistic case scenario with $\sim 50\,\%$ by modelling 20\,\% of the
galaxies as point masses. However, we consider such a compact mass
distribution at galaxy scales too contrived to be included in the
simulations.}. At lower redshifts ($z\lesssim 0.5$), the effects from
lensing are small and correcting for lensing is not likely to improve
the results.

Even though the fraction of SNe lines-of-sight passing through galaxy
cluster lenses is expected to be relatively low, these are potentially
important due to magnification bias and in surveys specifically aimed
at using cluster potentials as gravitational telescopes
\citep{gunngoo}. In those cases, individual modeling of the cluster
potentials using, e.g., weak and strong lensing of background galaxies
is needed to correct the observed magnitudes for the lensing
magnification. Alternatively, one can choose to discard SNe background
to galaxy clusters with very uncertain matter distributions when
estimating cosmological parameters. 

Our method also takes into account gravitational lensing from large
scale dark matter structures such as filaments and walls as long as
the matter density is dominated by the individual galaxy size
halos. If we very conservatively assume that large scale structures
are completely uncorrelated with the luminous matter, we expect the
lensing magnification contribution to be less than 2\,\% on scales
larger than 5 arcminutes (for a source redshift of unity)
\citep{cooray2005}.

For a given galaxy catalog, the computed magnifications do not
depend strongly on the cosmological parameters used. However, since
the formation of matter structure is a function of cosmology, it
should in principle be possible to determine the cosmology from the
observed distribution of standard candle luminosities. In this case,
the use of magnifications of e.g.~SNIa would be similar to using shear of 
background galaxies as in weak lensing studies. Such a
study would require large statistics of very well observed SNIa and
will probably have to await future dedicated missions such as the
proposed SNAP satellite~\footnote{\texttt{http://snap.lbl.gov}}.

For current high-$z$ SNIa observations the concern is to correct for
the magnification and investigate the possibility for magnification
bias. Such a study for SNe in the GOODS fields is described in a
accompanying paper \citep{jacke} where magnification bias is shown to
be negligible but lensing for individual SNe can be estimated quite
robustly from foreground galaxy observations and thus be corrected
for. In fact, as long as the luminous and dark matter are not
anti-correlated, we would expect to be able to reduce the scatter in
the Hubble diagram by assuming that dark matter follows light. Thus,
we find that even though the exact relation between luminous and dark
matter is uncertain, correcting for gravitational lensing using
observed galaxy properties should be harmless at the worst and very
useful at the best.

The authors would like to thank Mariangela Bernardi for help with the
velocity dispersion data from the Sloan survey, Joakim Edsj\"o, Daniel Holz and
Saul Perlmutter for helpful discussions during the course of the
work. We are grateful to Swara Ravindranath for providing simulated galaxy
catalogs used to derive the recovery fraction of galaxy morphologies.
CG would like to thank the Swedish Research Council
for financial support.
AG is a Royal Swedish Academy Research
Fellow supported by grants from the Swedish Research Council, the Knut and Alice Wallenberg
Foundation and the G\"oran Gustafsson Foundation for Research
in Natural Sciences and Medicine.


\clearpage

\begin{table}
\begin{center}
 \begin{tabular}{lrrrr} 
 \hline\hline
\multicolumn{5}{c}{Equation: $y=kz+m$} \\
\hline
& \multicolumn{2}{c}{Uncorrected} & \multicolumn{2}{c}{Corrected} \\
& \multicolumn{1}{c}{$k$} & \multicolumn{1}{c}{$m$} &
\multicolumn{1}{c}{$k$} & \multicolumn{1}{c}{$m$} \\ 
\hline
68\,\% magn. & -0.038 & 0.005 & -0.020 & 0.008 \\
68\,\% demagn. & 0.060 & -0.009 & 0.016 & 0.006 \\
95\,\% magn. & -0.140 & -0.003 & -0.055 & 0.008 \\
95\,\% demagn. & 0.086 & -0.017 & 0.035 & 0.012 \\
\hline
 \end{tabular}
\caption{Fits to the confidence levels of Fig.~\ref{fig:wcrc}
  expressed in magnitudes. Note that these are only valid for $z\gtrsim
  0.25$. However, below this value the corrections are indeed very small.}
\label{tab:lines}
\end{center}
\end{table}

\clearpage

\clearpage

\clearpage

\clearpage

\clearpage

\clearpage

\clearpage

\end{document}